\documentclass[structabstract]{aa}  
\usepackage{amsmath}
\usepackage{graphicx}
\usepackage{txfonts}
\usepackage{natbib}
\usepackage{rotating}	
\usepackage{color}

\usepackage{multirow}
\usepackage{soul}
\usepackage{float}
\bibpunct{(}{)}{;}{a}{}{,} 
\begin{document}
   \title{The star formation rate cookbook at $1< z < 3$: \\Extinction-corrected relations for UV $\&$ [OII]$\lambda$3727 luminosities}

\author{M. Talia
\inst{1,2}
\and
A. Cimatti\inst{1} \and
L. Pozzetti\inst{2} \and
G. Rodighiero\inst{3} \and
C. Gruppioni\inst{2} \and
F. Pozzi\inst{1} \and
E. Daddi\inst{4} \and
C. Maraston\inst{5} \and
M. Mignoli\inst{2} \and
J. Kurk\inst{6} 
}
          
\offprints{M. Talia\\
\email{margherita.talia2@unibo.it}}

\institute{Dipartimento di Fisica e Astronomia, Universit\`a di Bologna,
Via Ranzani 1, I-40127, Bologna, Italy
\and 
INAF- Osservatorio Astronomico di Bologna,
Via Ranzani 1, I-40127, Bologna, Italy
\and 
Universit\`a di Padova, Dipartimento di Astronomia,
vicolo dell'Osservatorio 2, I-35122, Padova, Italy
\and
CEA-Saclay, DSM/Irfu/Service d'Astrophysique, 91191
Gif-sur-Yvette Cedex, France
\and
Institute of Cosmology and Gravitation, Dennis Sciama Building, 
Burnaby Road, Portsmouth PO1 3FX, UK
\and
Max Planck Institut f\"ur extraterrestrische Physik,
Postfach 1312, 85741 Garching bei M\"unchen, Germany
}

   \date{}
 
  \abstract
   {}
   {In this paper we use a well-controlled spectroscopic sample of galaxies at $1{<}z{<}3$ drawn from the \emph{Galaxy Mass Assembly ultra-deep Spectroscopic Survey} (GMASS) to study different star formation rate (SFR) estimators. In particular, we use infrared (IR) data to derive empirical calibrations to correct ultraviolet (UV) and [OII]$\lambda$3727 luminosities for dust extinction and dust-corrected estimates of SFR. 
   }
   {We selected 286 star-forming galaxies with spectroscopic redshift $1{<}z{<}3$. In order to have a homogeneous wavelength coverage in the spectra, the sample was divided into two sub-groups: galaxies at $1{<}z{<}1.6$ whose spectra cover the rest-frame range ${\sim}2700{-}4300\AA$, where the [OII]$\lambda$3727 emission line can be observed, and galaxies at $1.6{<}z{<}3$ whose spectra cover the range ${\sim}1100{-}2800\AA$. 
   In the selection procedure we fully exploit the available spectroscopic information. In particular, on the basis of three continuum indices, we are able to identify and exclude from the sample galaxies in which old stellar populations might bring a non-negligible contribution to IR luminosity ($L_{IR}$) and continuum reddening. Using Spitzer-MIPS and Herschel-PACS data we derive $L_{IR}$ for two-thirds of our sample. The $L_{IR}/L_{UV}$ ratio is used as a probe of effective attenuation ($A_{IRX}$) to search for correlations with continuum and spectroscopic features in order to derive empirical calibrations to correct UV and [OII]$\lambda$3727 luminosities for dust extinction.
   }
   {Through the analyses of the correlations between different dust attenuation probes, a set of relations is provided that allows the recovery of the total unattenuated SFR for star-forming galaxies at $1{<}z{<}3$ using UV and [OII]$\lambda$3727 luminosities. 
   
   The relation between $A_{IRX}$ and UV continuum slope ($\beta$) was tested for our sample and found to be broadly consistent with the literature results at the same redshift, though with a larger dispersion with respect to UV-selected samples.
   
   We find a correlation between the rest-frame equivalent width of the [OII]$\lambda3727$ line and $\beta$, which is the main result of this work. We therefore propose the rest-frame equivalent width of the [OII]$\lambda3727$ line as a dust attenuation probe and calibrate it through $A_{IRX}$, though the assumption of a reddening curve is still needed to derive the actual attenuation towards the [OII]$\lambda3727$ line ($A_{[OII]}$). We tested the issue of differential attenuation towards stellar continuum and nebular emission: our results are in line with the traditional prescription of extra attenuation towards nebular lines. 
   
   Finally, we use our set of cross-calibrated SFR estimates to look at the relation between SFR and stellar mass. The galaxies in our sample show a close linear relation ($\sigma{=}0.3$ dex) at all redshifts with a slope ${\sim}0.7-0.8$, which confirms several previous results.
   }
   {}


   \titlerunning{The star formation rate cookbook at $1< z < 3$}
   \maketitle
%

\section{Introduction}\label{sec:Introduction}

In the past decades, great effort has been devoted to the study of the Universe in the redshift range $1{<}z{<}3$. Several studies have shown that this is the epoch when a substantial fraction of galaxy mass assembly took place, and when there is a peak in the evolution of the star formation rate (SFR) density through cosmic time \citep{lilly1996, madau1996, dickinson2003, hopkins2006, daddi2007}.
In this epoch a critical transformation phase is believed to have occurred, revealed by the observed changes in the colour-mass plane where a significant fraction of galaxies moves from the blue cloud of active star formation to the red sequence inhabited by spheroidal galaxies with weak or suppressed star formation at $z{<}1$ \citep[e.g.][]{cassata2008, cimatti2013}.

A key element in the study of galaxy evolution is obviously the availability of large samples of galaxies. Spectroscopic redshift surveys play a crucial role as they provide samples with confirmed redshifts. Though photometric redshift surveys have now reached a high level of accuracy, the number of catastrophic failures, even if low at a few percent (Ilbert et al. 2013), could still produce large unknowns. Moreover, spectroscopy provides information that is not accessible by broad-band photometry, e.g. emission line fluxes and absorption features that are fundamental to gaining insight on gas properties and therefore dust extinction, star formation, and feedback mechanisms.

Astronomical observation plans are actually going in the direction of collecting spectroscopic information for increasingly large samples of galaxies, as confirmed by the number of spectroscopic surveys that have been carried out in recent years or are planned for the near future (e.g. K20 \citep{mignoli2005}, VVDS \citep{lefevre2005}, Deep2 \citep{willmer2006}, zCOSMOS \citep{lilly2007}, ESO-GOODS \citep{vanzella2008, popesso2009, balestra2010}, PRIMUS \citep{coil2011}, GMASS \citep{kurk2012}, DEIMOS/Keck (Kartaltepe et al., in prep.; Capak et al., in prep.), VUDS \citep{lefevre2014}, MOSDEF \citep{kriek2015}, VANDELS\footnote[1]{$http{:}//vandels.inaf.it/$}, BigBOSS \citep{schlegel2012}, Euclid \citep{laureijs2011}, WFIRST \citep{spergel2013}). 
The large number of high-redshift galaxy spectra that is rapidly becoming available inspires the need to test existing analysis methods and calibrations, and to create new ones, to correctly interpret the wealth of spectroscopic information in the frame of understanding the main processes that regulate galaxy evolution.

The rate at which a galaxy produces stars is a critical ingredient to the investigation of galaxy evolution. Though other properties like stellar mass can be quite robustly estimated from SED fitting to photometric data, there is still room for improvement in our ability to estimate the star formation rate (SFR) at high redshift because of the uncertainties in the assumptions made in translating integrated fluxes and colours to an estimate of the SFR \citep{madau2014, pannella2014}. 

The rest-frame UV continuum light emitted by young massive stars and the H$\alpha$ optical emission line are very good and primary tracers of star formation \citep[e.g.][]{kennicutt1998, kennicutt2012}. At high redshift, where H$\alpha$ is less easily accessible, the [OII]$\lambda$3727 emission line can be a fair alternative. However, these primary and secondary tracers are known to be affected by the presence of dust that absorbs the flux emitted at UV and optical wavelengths, and re-emits it at longer wavelengths in the infrared (IR) regime. 
The use of dust-unbiased tracers like mid- and far-IR (FIR) or radio continuum would be preferable, but it is still limited at high redshift due to sensitivity limits. Giant strides have been made in this respect with the advent of the \emph{Spitzer} and \emph{Herschel} telescopes in the last decade \citep[see e.g.][]{lutz2011, rodighiero2010, elbaz2011, nordon2013, oteo2014}, but the low-SFR regimes at high redshift can still only be accessed by means of stacking \citep{rodighiero2014, pannella2014}. 
In order to study star-forming galaxies in a large range of SFR, UV and line emission luminosities are still the best choice for cosmologically relevant galaxy samples, and the study of reliable corrections for dust attenuation continues to be a crucial topic. 

Dust attenuation affecting emission line luminosities can be derived by measuring the Balmer decrement in galaxy spectra  \citep{brinchmann2004, garn2010}, but this information is still rarely available at high redshift. Indirect ways to correct for dust extinction are often employed, such as the comparison with other estimates of SFR like the value derived from SED fitting to broad-band photometry \citep{forsterschreiber2009} under the assumption of some attenuation curves \citep[e.g.][]{calzetti2000}.

The dust correction of UV light commonly relies on the local correlation between the slope of the UV continuum and dust attenuation \citep{meurer1999, calzetti2000, daddi2004, overzier2011, takeuchi2012}, though the general validity of such correlation, especially at high redshift, is still under debate \citep{calzetti2001, boissier2007, seibert2007, reddy2010, reddy2012, reddy2015, overzier2011, takeuchi2012, buat2012, heinis2013, nordon2013, castellano2014, oteo2013, oteo2014, hathi2015}.

Though the data coming from recent IR surveys are not sufficient to map the population of SFGs down to low-SFR regimes, they can be effectively used to test the existing calibrations of dust extinction correction recipes, or to derive new ones. This can be made under the assumption that the IR emission depends only on the absorption of the flux emitted by the young stars responsible of UV light. However, there might be a non-negligible contribution of old stars to the dust heating that grows with decreasing sSFR  \citep{dacunha2008, kennicutt2012, arnouts2013, utomo2014}, leading to an overcorrection of UV light when IR luminosity ($L_{IR}$) is used as dust correction probe. 
A way to overcome this problem is to use spectroscopic samples, where continuum indices may be computed that indicate the possible presence of old stellar populations \citep{bruzual1983, daddi2005, cimatti2008}.

In this paper, different indicators of star formation for a sample of $1<z<3$ galaxies will be analysed. We concentrate on a well-controlled spectroscopic sample, rich of ancillary panchromatic photometric data, including IR from \emph{Spitzer} and \emph{Herschel}, in order to focus on the information that can be obtained from spectra. We will use the IR-derived SFR as a benchmark against which compare other estimates at shorter wavelengths, namely UV and [OII]$\lambda3727$ luminosities, and to derive new (or update existing) calibrations of dust extinction correction recipes. This way we are able to build a set of self-consistent recipes to derive the total un-extincted SFR, which are collected in a Table at the end of the paper. 

Throughout this paper, we adopt $H_{0}{=}70$ km/s/Mpc, $\Omega_{m}{=}0.3$, $\Omega_{\Lambda}{=}0.7$, give magnitudes in AB photometric system, and assume a \citet{kroupa2001} initial mass function. All linear fits, unless differently stated in the text, will be ordinary least squares (OLS) regressions of \emph{Y} on \emph{X} (Y$|$X) \citep{bevington1969, isobe1990}. Finally, we will indicate with the suffix "0" all un-extincted quantities, i.e. quantities corrected for dust extinction.

\section{The multi-wavelength dataset}\label{sec:The sample of star-forming galaxies}

The main ingredient of the present study is a well-controlled spectroscopic sample at $1<z<3$ selected from the photometric catalogue of the \emph{Galaxy Mass Assembly ultra-deep Spectroscopic Survey} (GMASS). 

\subsection{The GMASS survey}\label{sec:The GMASS survey}

The GMASS\footnote[2]{$http{:}//www.mpe.mpg.de/{\sim}kurk/gmass/$} survey \citep{kurk2012} is an ESO VLT large program project based on data acquired using the FOcal Reducer and low dispersion Spectrograph (FORS2). 
The project's main science driver is to use ultra-deep optical spectroscopy to measure the physical properties of galaxies at redshifts $1.5{<}\emph{z}{<}3$. 
The \emph{GMASS photometric catalogue} is a pure magnitude limited catalogue from the GOODS-South public image in the IRAC band at $4.5 \mu m$: $\emph{m}_{4.5}{<}23.0$ (AB system). In the chosen redshift range this selection is most sensitive to stellar mass. In particular, the limiting mass sensitivities are $log(M/M_{\odot}) \sim$ 9.8, 10.1, and 10.5 for $z =$ 1.4, 2, and 3, respectively. 
The GMASS photometric catalogue gathers information from U band to IRAC 8.0 $\mu$m band of 1277 objects. 
	\begin{figure}[h!]
	\centering
	\includegraphics[scale=0.43]{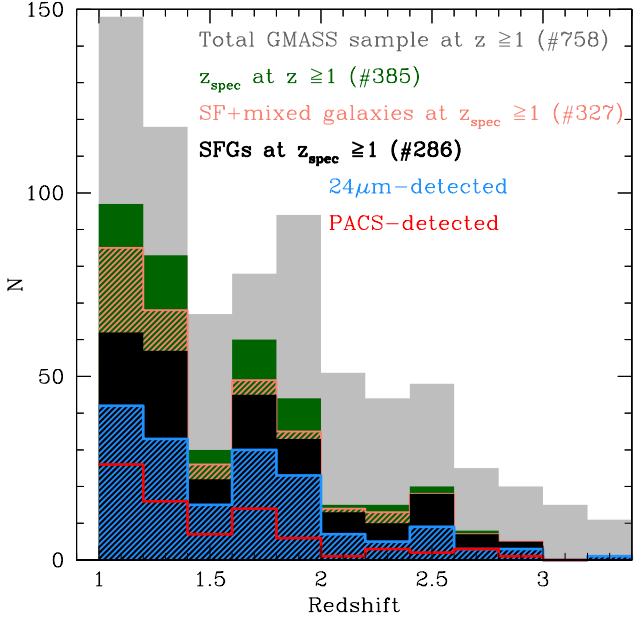}
	\caption{Redshift distribution of total GMASS sample at $z \geq 1$ with either spectroscopic or photometric redshift (grey); GMASS sample at $z_{spec} \geq 1$ with reliable spectroscopic redshift (green); preliminary sample of 327 SF+\emph{mixed} galaxies (salmon); final sample of 286 SFGs analysed in the paper (black); $24 \mu m$-detected SFGs (blue); PACS-detected SFGs (red).}
	\label{hist_z1}
	\end{figure}

\subsection{Spectroscopic data}\label{sec:Publicly available spectroscopic data in the CDF-S}

From the GMASS photometric catalogue, 170 objects were selected as main targets to be observed (250 objects including fillers) during 145h, and a secure spectroscopic redshift could be determined for 131 of them. The spectral resolution of GMASS spectra is $R = \lambda/\Delta\lambda \sim 600$.
To extend our analysis, we collected public spectra from other spectroscopic surveys for the galaxies in the GMASS photometric catalogue with no GMASS spectrum. In particular, we searched for counterparts of GMASS galaxies from the following surveys: the ESO-GOODS/FORS2 v3.0 \citep{vanzella2008} and ESO-GOODS/VIMOS 2.0 \citep{popesso2009, balestra2010}, the VVDS v1.0 \citep{lefevre2005}, and the K20 \citep{mignoli2005}. We refer the reader to the cited papers for more information about the various surveys.

A total of 385 galaxies with a reliable spectroscopic redshift $z > 1$, coming from any one of the cited spectroscopic surveys, was selected (see Fig. \ref{hist_z1}). 

\subsection{Mid- and far-IR photometry}\label{sec:Herschel-PACS combined PEP/GOODS-H observations}

The GOODS-South field has been observed with the Photodetector Array Camera and Spectrometer (PACS; \citet{poglitsch2010}) on board the Herschel Space Observatory\footnote[3]{Herschel is an ESA space observatory with science instruments provided by European-led Principal Investigator consortia and with important participation from NASA \citep{pilbratt2010}.} as part of two projects: the \emph{PACS Evolutionary Probe} (PEP; \citet{lutz2011}) and the \emph{GOODS-Herschel} \citep{elbaz2011} programmes. The publicly released PACS catalogue produced using as priors the source positions expected on the basis of a deep Spitzer-MIPS 24 $\mu m$ catalogue \citep{magnelli2011} was used in this work. 
We refer the reader to \citet{magnelli2013} for all the details about data reduction and the construction of images and catalogues. 
The public catalogue presented in \citet{magnelli2013} has a 24 $\mu m$ flux cut at 20$\mu Jy$. We extended it to fainter fluxes with a 24 $\mu m$ source catalogue from Daddi et al. (in prep.) that was created, similarly to the \citet{magnelli2011} one, using a PSF fitting technique at the 3.6 $\mu m$ source positions as priors.

GMASS coordinates of the 385 galaxies at $1{<}z_{spec}{<}3$ were matched to the sources listed in the 24 $\mu m$ source catalogues. Counterparts were searched within an angular separation of 1.4". After some tests, this angular separation was chosen as the best compromise not to lose real counterparts, while minimizing the selection of false pairs. 
236 24 $\mu m$ counterparts were found, 121 of which have at least one PACS detection.

\section{Selection of the star-forming galaxies (SFGs)}\label{sec:The final SFG sample}

Since the aim of this work is to study the star formation in star-forming galaxies (SFGs) at $1{<}z{<}3$, the catalogue assembled of 385 galaxies at $z_{spec}{\geq}1$ had to be cleaned of quiescent objects and active galactic nuclei (AGNs). 

\begin{figure*}[t!]
	\centering
	\includegraphics[scale=0.88, trim=0mm 0mm 0mm 52mm, clip=true]{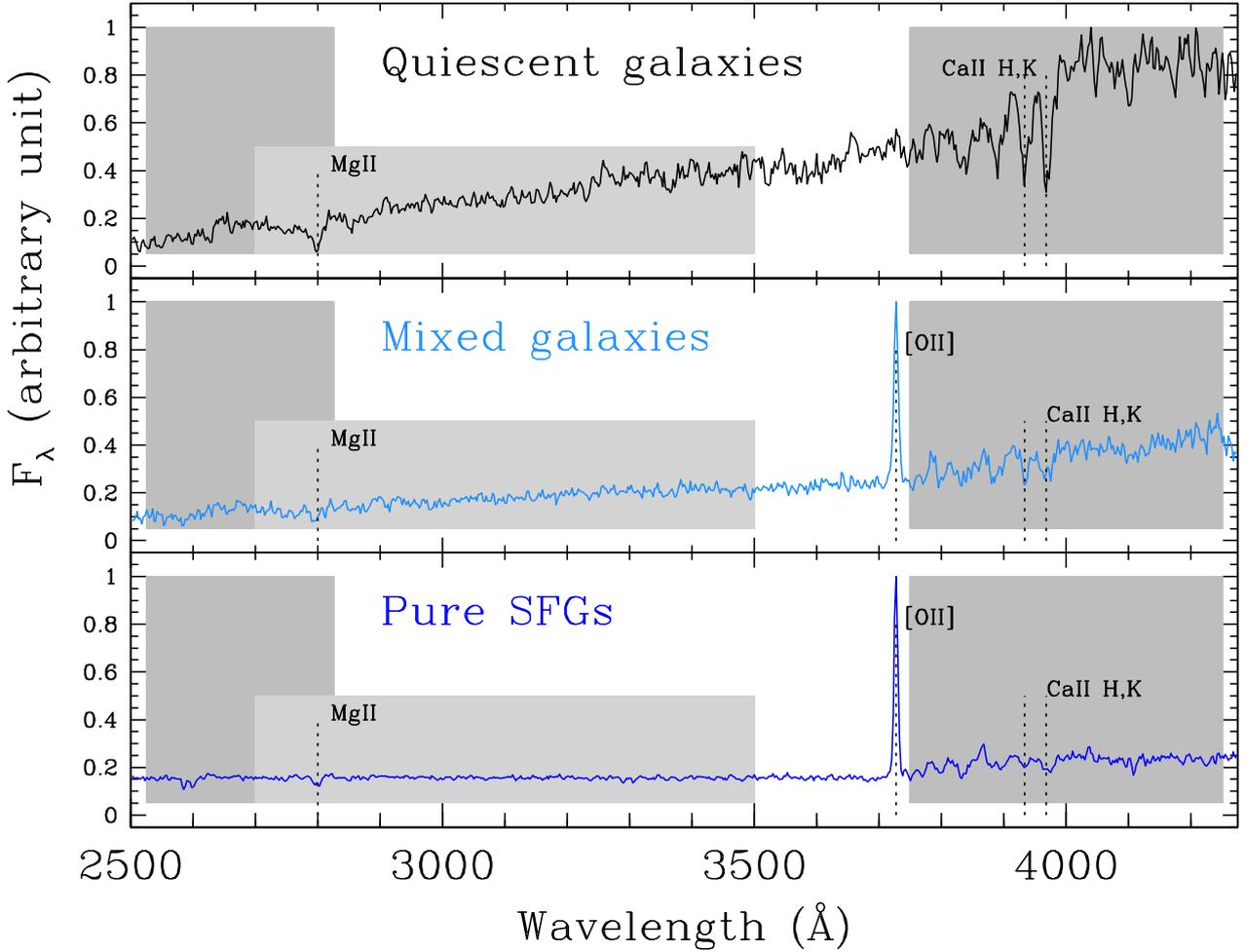}
	\caption{From top to bottom: 	a) average spectrum of 33 quiescent galaxies in the GMASS sample; 
		 			b) average spectrum of \emph{mixed} galaxies in the \emph{[OII] sample}, at $1{<}z{<}1.6$;
					c) average spectrum of not-\emph{mixed} SFGs in the \emph{[OII] sample}, at $1{<}z{<}1.6$. 
		\emph{Mixed} galaxies have high values of the three continuum indices indicated by grey shaded regions: MgUV ($2600-2900\AA$) \citep{daddi2005}; $C(29-33)$ ($2900-3300\AA$) \citep{cimatti2008}; $D4000$ break ($3750-4250\AA$) \citep{bruzual1983}. See the text and Fig. \ref{index_hist} for more details.}
	\label{spectra1}
	\end{figure*}

On the basis of the spectroscopic features, 33 quiescent galaxies were identified and excluded from the sample. Their average spectrum (Fig. \ref{spectra1}) shows the characteristic features of the rest-frame UV spectrum of old and passive stellar populations: the very red continuum, characterized by prominent breaks (at 2640$\AA$, 2900$\AA$, 4000$\AA$) and rich of metal absorptions (e.g. CaII H$\&$K doublet) \citep{cimatti2008, mignoli2005}, and the weakness or complete absence of [OII]$\lambda$3727 emission line. We will come back on the shape of the continuum of old stellar populations later in this section.

A combination of spectroscopic features and X-ray information was instead used to identify AGN hosts. In particular, we excluded galaxies showing typical broad and/or narrow emission lines (e.g. CIV$\lambda$1549$\AA$, MgII$\lambda$2800). We also excluded galaxies with an X-ray detection from the Chandra 4Ms catalogue \citep{xue2011}, and absorption-corrected rest-frame 0.5-8 kev luminosity higher than $L_{x} = 3 x 10^{42}$ erg/s, following the criteria for AGN identification given by \citet{xue2011}. This selection made us discard 25 more galaxies from the sample.
The objects with an X-ray detection, but $L_{x} < 3 x 10^{42}$ erg/s were instead kept in the sample. 

\begin{figure}[b!]
	\centering
	\includegraphics[scale=0.43]{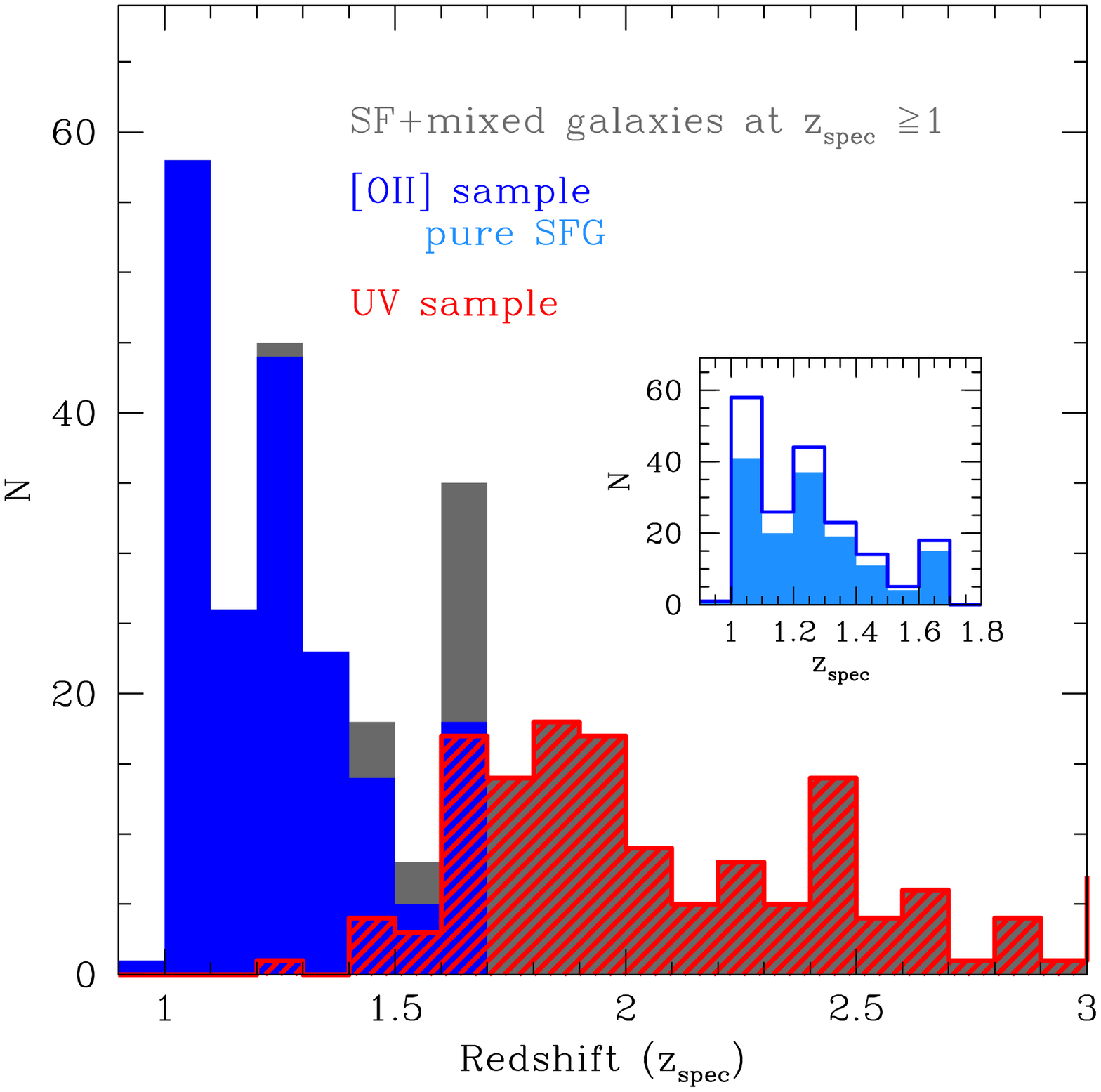}
	\caption{Redsfhit distribution. Big plot: the grey histogram represents the preliminary selection of 327 SF+\emph{mixed} galaxies; the blue histogram represents the \emph{[OII] sample} (i.e. spectra covering the wavelength range $\sim2700-4300\AA$); the red histogram represents the \emph{UV sample} (i.e. spectra covering the wavelength range $\sim1100-2800\AA$). Little inset: zoom of the \emph{[OII] sample}: empty blue histogram represents the total \emph{[OII] sample}, while light blue shaded histogram represents only not-\emph{mixed} SFGs in the \emph{[OII] sample}.}
	\label{hist_z2}
	\end{figure}

The remaining, at this stage, 327 galaxies showed spectroscopic features typical of SFGs and lack of AGN evidence from the X-rays. 
UV spectroscopic features identifying a SFG are the presence of strong nebular emission lines, such as [OII]$\lambda$3727, and/or a blue UV continuum. Not all the features can be detected in all the spectra because, given the wide range of redshifts covered by our sample, the galaxy spectra sample different rest-frame wavelength ranges. In particular, we can roughly divide the sample into two sub-groups: galaxies at $1{<}z{<}1.6$ that cover the range $\sim2700-4300\AA$ (\emph{[OII] sample}), where the [OII]$\lambda$3727 emission line can be observed; and galaxies at $1.6{<}z{<}3$ that cover the range $\sim1100-2800\AA$ (\emph{UV sample}), in whose spectra strong inter-stellar medium (ISM) absorption lines can be detected (see Fig. \ref{hist_z2}). 

\subsection{Spotting the presence of old stellar populations in SFGs}\label{sec:Mixed galaxies}

In this work we decided to use $L_{IR}$ as benchmark SFR indicator, exploiting the recent HERSCHEL data in combination with deep spectroscopy. $L_{IR}$ is the bolometric luminosity coming from dust emission and can be directly converted into the rate of obscured star formation \citep{kennicutt1998, kennicutt2012}. The validity of such conversion relies on the assumption that only young star populations (the ones responsible for the UV luminosity emission) heat the dust that than re-emits in the IR. However, there might be a non-negligible contribution of old stars to the dust heating that grows with decreasing sSFR  \citep{dacunha2008, kennicutt2012, arnouts2013, utomo2014}, leading to an overestimate of the true SFR when using $L_{IR}$ as sole estimator.

Thanks to the spectroscopic information we are able to overcome this issue. In the wavelength range covered by our spectra three continuum indices may be computed that indicate the possible presence of old stellar populations. These indices are: MgUV, which is a feature produced by the combination of the strongest breaks and absorptions at $2600-2900\AA$ \citep{daddi2005}; $C(29-33)$, which is a colour index of the UV continuum defined at $2900-3300\AA$ \citep{cimatti2008}; and the $D4000$ break \citep{bruzual1983}. We refer the reader to the cited papers for the definitions of the indices. The range over which each parameter is defined is indicated in Fig. \ref{spectra1}.

Typically, early-type galaxies have $MgUV>1.2$, $C(29-33)>0.6$, $D4000>1.6$ \citep{daddi2005, cimatti2008, mignoli2005, kauffmann2003, hathi2009}. We measured the indices on the spectra of the 327 galaxies resulting from the previous selection in order to search for evidence of the possible presence of old stellar populations along the young ones. \\ 

\noindent \underline{\emph{\textbf{[OII] sample}}}

\begin{figure}[h!]
	\centering
	\includegraphics[scale=0.43]{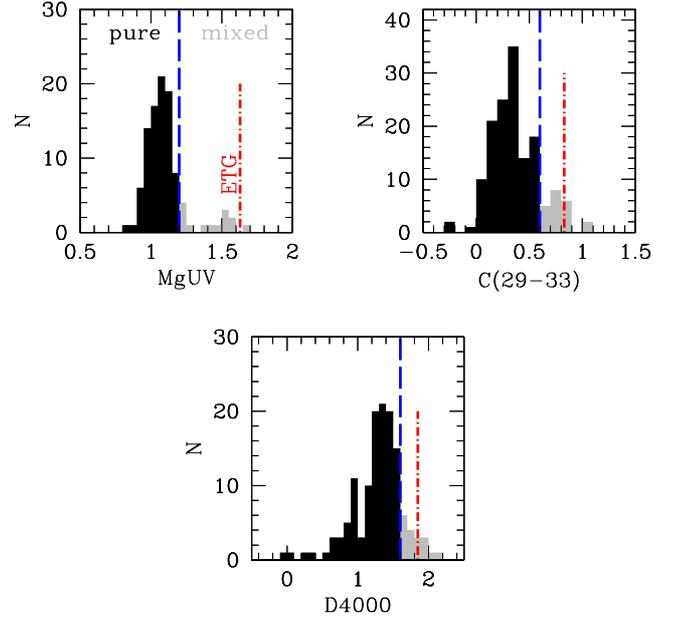}
	\caption{Continuum indices distribution over the \emph{[OII] sample}: MgUV, C(29-33), D4000. In each plot the blue vertical line represents the threshold that separates SFGs from \emph{mixed} galaxies (see the text for more details). The red lines mark the average value of each index measured on the GMASS quiescent galaxies at $z_{spec}\geq 1$, for reference.}
	\label{index_hist}
	\end{figure}

\noindent Given the different rest-frame wavelength coverage of the spectra, not all indices are available for all galaxies. 
The spectra of the galaxies in the \emph{[OII] sample} always have at least one index (often two) in their range. We find that $22\%$ of the \emph{[OII] sample} have a continuum suggesting the presence of an old stellar population that might contribute to dust heating and to the reddening of the UV continuum.
The parameters distributions are shown in Fig. \ref{index_hist}, and the mean value of each index measured on the 33 quiescent galaxies is also indicated as reference. 
To make a distinction, we will call the objects whose indices are above the threshold \emph{mixed} galaxies (because they show both the [OII]$\lambda$3727 line and a red continuum), while the objects whose indices are below the thresholds will be referred to simply as SFGs in the rest of the paper. The relative redshift distributions of the two classes are shown in Fig. \ref{hist_z2}.
The average spectrum of \emph{mixed} galaxies, presented in Fig. \ref{spectra1}, though showing the [OII]$~\lambda 3727$ emission line, has a continuum shape more similar to that of quiescent galaxies, whose average spectrum is shown as reference. 
Only SFGs (i.e. not-\emph{mixed}) are used in our analysis. \\

\noindent \underline{\emph{\textbf{UV sample}}}

\noindent In the \emph{UV sample} only the MgUV index is available in the spectrum of $\sim45\%$ of the sample. For the remaining $\sim55\%$ of the \emph{UV sample} no index is available. When in range, the MgUV index is always $<1.2$, within the errors. Moreover, all the galaxies in the \emph{UV sample} (with and without measurable MgUV index) have the bluest colour in the \emph{(NUV-r) vs. (r-K)} colour-colour plot, implying that they have the highest sSFR \citep{arnouts2013}. This means that in these galaxies a possible contribution of old stellar populations to dust heating should be negligible \citep[see also][]{utomo2014}.
Therefore we classify all the galaxies in the \emph{UV sample} as not-\emph{mixed} SFGs, and include them all in the analysis.

\subsection{Summary of the final sample selection}\label{sec:The final SFG sample}
	\begin{table*}[t!]
	\centering             
	\caption{Summary table of the general properties (spectral classes, X-ray and IR detections) of the \emph{GMASS} spectroscopic sample at $z_{spec}\geq1$.}
	\label{general}
	\begin{tabular}{|l|l|c c c| c c c| c c|}  
	\hline\hline
	  \multicolumn{10}{|c|}{}\\
	  \multicolumn{10}{|c|}{\emph{GMASS} galaxies with spectroscopic data and reliable spectroscopic redshift ($z_{spec}\geq1$)} \\
	  \multicolumn{10}{|c|}{}\\
	\hline\hline  
	  \multicolumn{2}{|c|}{\emph{[OII] sample} 219 objs.} & quiescent & SFG & AGN & $L_{X}{>}3{\times}10^{42}$erg/s & $L_{X}{<}3{\times}10^{42}$erg/s & NO $L_{X}$ & $24\mu$ m           & PACS\tablefootmark{a}\\
	\hline
		  & quiescent                            &  21       &  0  &  0  &   0                           &   4                           &   17    &       6                  & 4\\
        Spectral & SFG\tablefootmark{b}     	     &           &192  & 0   &  4                            &   15                          &   173    &     130                  & 72\\
	class    & AGN                                  &           &     &  6  &   2                           &      0                        &   4      &        3                 & 1\\
	\hline
	         & $L_{X}{>}3{\times}10^{42}erg/s$        &           &     &     &   6                           &   0                           &   0      &       6                  & 3\\
	X-ray    & $L_{X}{<}3{\times}10^{42}erg/s$        &           &     &     &                               &    19                         &    0     &       15                 & 11\\
	         & NO $L_{X}$                  	       &           &     &     &                               &                               &   194    &       118                & 63\\
	\hline
	IR       & $24\mu$ m                              &           &     &     &                               &                               &          &      139                 & 77\\
	data     & PACS\tablefootmark{a}                &           &     &     &                               &                               &          &                          & 77\\
	\hline\hline
	  \multicolumn{2}{|c|}{\emph{UV sample} 166 objs.}    & quiescent & SFG & AGN & $L_{X}{>}3{\times}10^{42}$erg/s & $L_{X}{<}3{\times}10^{42}$erg/s & NO $L_{X}$ & $24\mu$ m            & PACS\tablefootmark{a}\\
	\hline
		 & quiescent                            &    12       &  0   &  0   &               1                &            1                   &    10      &      4                    & 3\\
        Spectral & SFG\tablefootmark{b}     	        &           &  148   & 0    &               9                &           3                    &   136       &      88                    &  36\\
	class    & AGN                                  &           &     &  6   &               5                &             0                  &    1      &      5                    & 5\\
	\hline
	         & $L_{X}{>}3{\times}10^{42}$erg/s        &           &     &     &               15                &             0                  &   0       &       12                   & 10\\
	X-ray    & $L_{X}{<}3{\times}10^{42}$erg/s        &           &     &     &                               &            4                   &    0      &       2                   & 1\\
	         & NO $L_{X}$                  	       &           &     &     &                               &                               &   147       &        83                  & 33\\
	\hline
	IR       & $24\mu$ m                              &           &     &     &                               &                               &          &        97                  & 44\\
	data     & PACS\tablefootmark{a}                &           &     &     &                               &                               &          &                          & 44\\
	\hline\hline	
	\end{tabular}
	\tablefoot{
	\tablefoottext{a}Detection in at least one of the three PACS bands.
	\tablefoottext{b}Preliminary selection of SFGs. It includes both not-\emph{mixed} SFGs and \emph{mixed} galaxies. See the text for more details (Sec. \ref{sec:Mixed galaxies}).
	 }
	\end{table*}
	\begin{table*}[t!]
	\centering             
	\caption{Summary table of the general properties (spectral classification and continuum indices, X-ray, and IR detections) of SFGs and \emph{mixed} galaxies (Tot. 327 galaxies).}
	\label{sel}
	\begin{tabular}{|c|c|c|c|c||c|c|c||c|c|c c c|}  
	\hline\hline
	                                     & \multirow{2}{*}{z}           & Spec.    & $L_{X}$                  & Tot.   & \multirow{2}{*}{MgUV}\tablefootmark{a}  & \multirow{2}{*}{C(29-33)}\tablefootmark{a} & \multirow{2}{*}{D4000}\tablefootmark{a} & \multirow{2}{*}{no IR} & MIPS    &         & PACS     &         \\
	                                     &                              & class    & (erg/s)                & num.   &                                         &                                            &                                         &                        & $24\mu$ m & $70\mu$ m & $100\mu$ m & $160\mu$ m\\
	\hline\hline
	\multirow{2}{*}{\emph{UV sample}}    & \multirow{2}{*}{$1.6{-}3.0$} & pure SFG & ${<}3{\times}10^{42}$  & 3      & \multirow{2}{*}{$1.07$}                  & \multirow{2}{*}{${-}$}                     & \multirow{2}{*}{${-}$}                  & 1                      &  2      &  0      &  0        &  1      \\
	                                     &                              & pure SFG & NO                     & 136    &                                         &                                            &                                         & 58                     &  78     &  4      &  25       &  19     \\
	\hline
	\multirow{4}{*}{\emph{[OII] sample}} & \multirow{4}{*}{$1.0{-}1.6$} & pure SFG & ${<}3{\times}10^{42}$  &  9     & \multirow{2}{*}{$1.06$}                 & \multirow{2}{*}{$0.31$}                    & \multirow{2}{*}{$1.30$}                 &  0                     &   9     &  2      &   6       &   5     \\
	                                     &                              & pure SFG & NO                     &  138   &                                         &                                            &                                         &  50                    &   88    &   11    &   42      &   37    \\
	                                     &                              & mixed    & ${<}3{\times}10^{42}$  &  6     & \multirow{2}{*}{$1.49$}                 & \multirow{2}{*}{$0.74$}                    & \multirow{2}{*}{$1.80$}                 &   2                    &   4     &   2     &   4       &   4     \\
	                                     &                              & mixed    & NO                     &  35    &                                         &                                            &                                         &   10                   &    25   &   2     &   12      &   9     \\		
	\hline\hline
	\end{tabular}
	\tablefoot{
	\tablefoottext{a}Median values over the sub-sample. The distribution of each parameter in the \emph{[OII] sample} is shown in Fig. \ref{index_hist}.
	 } 
	\end{table*}

From the magnitude limited GMASS photometric catalogue we preliminary selected 327 galaxies with a secure spectroscopic redshift $1{<}z{<}3$, typical spectroscopic features associated to star formation (presence of [OII]$\lambda$3727 emission line and/or a blue continuum), and lack of evidence of AGN presence from the X-rays. 
We summarize the main properties of the parent spectroscopic sample in Table \ref{general}, and give additional information on the preliminary sample of 327 galaxies in Table \ref{sel}. 
A deeper spectroscopic analysis of the preliminary selection of galaxies revealed the possible presence of old stellar populations in $\sim12\%$ of them. These 41 \emph{mixed} galaxies will be excluded from the analysis because of the possible contribution of old stars to dust heating and continuum reddening. The 286 SFGs will be used to study the relations between different estimators of SFR. In the rest of the paper, when talking about SFGs we will refer to the not-\emph{mixed} ones. 
In Table \ref{spec_survey} we report some additional spectroscopic information about the 286 SFGs.
	\begin{figure*}[t!]
	\centering
	\includegraphics[scale=0.78]{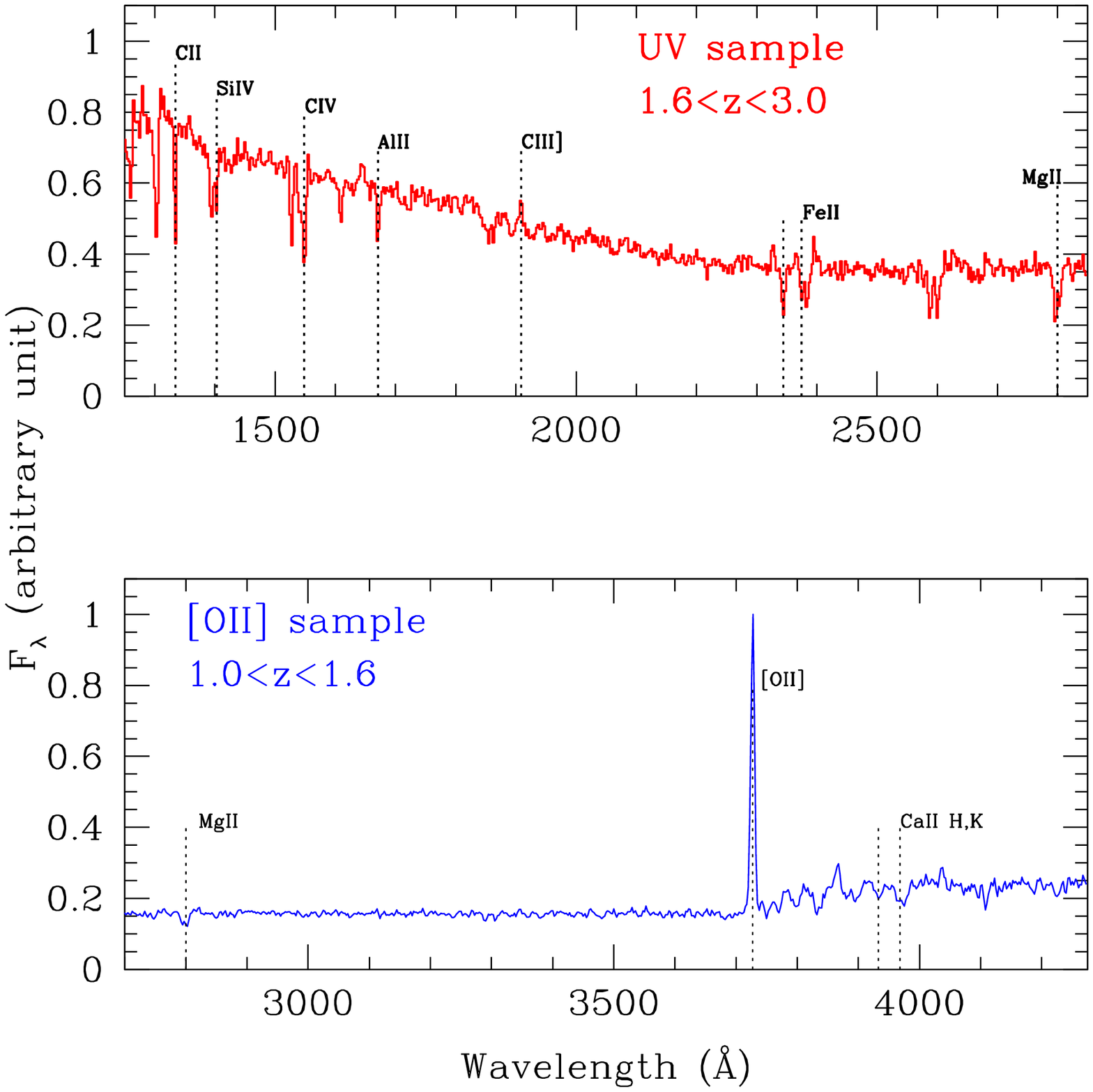}
	\caption{Top: average spectrum of SFGs in the \emph{UV sample}, at $1.6{<}z{<}3$; bottom: average spectrum of SFGs in the \emph{[OII] sample}, at $1{<}z{<}1.6$.}
	\label{spectra2}
	\end{figure*}

On the basis of the wavelength coverage of the spectra, the sample can be roughly divided into two sub-groups: galaxies at $1{<}z{<}1.6$ whose spectra cover the range $\sim2700-4300\AA$, and galaxies at $1.6{<}z{<}3$ whose spectra cover the range $\sim1100-2800\AA$. We will refer to these groups, respectively, as the \emph{[OII] sample} and the \emph{UV sample}.
In Fig. \ref{spectra2} the average spectra of the galaxies of the two sub-samples are shown.

\subsection{Comparison to the parent sample}\label{sec:Comparison to the parent sample}
	\begin{figure*}[t!]
	\centering
	\includegraphics[scale=0.43]{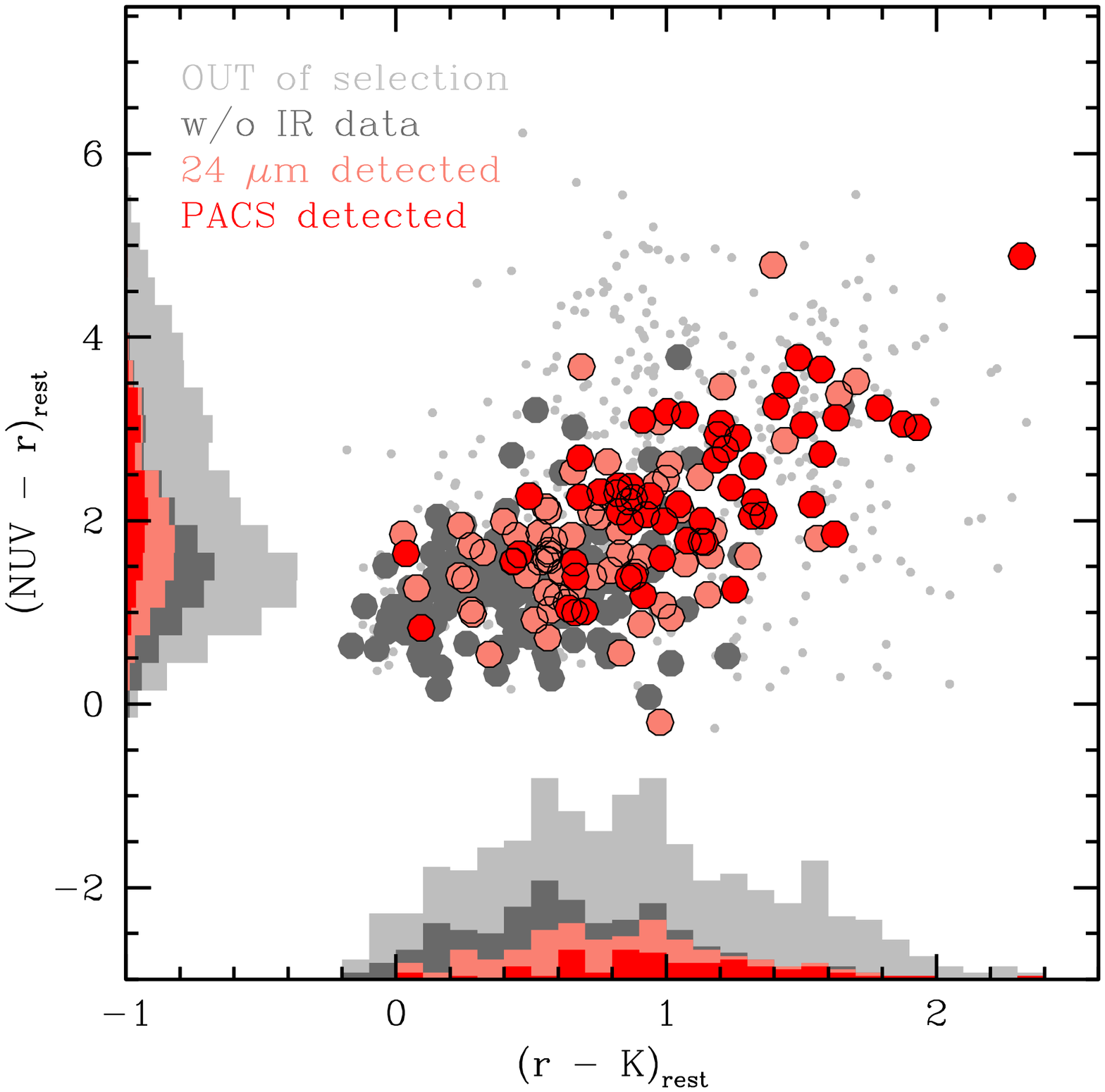}
	\includegraphics[scale=0.43]{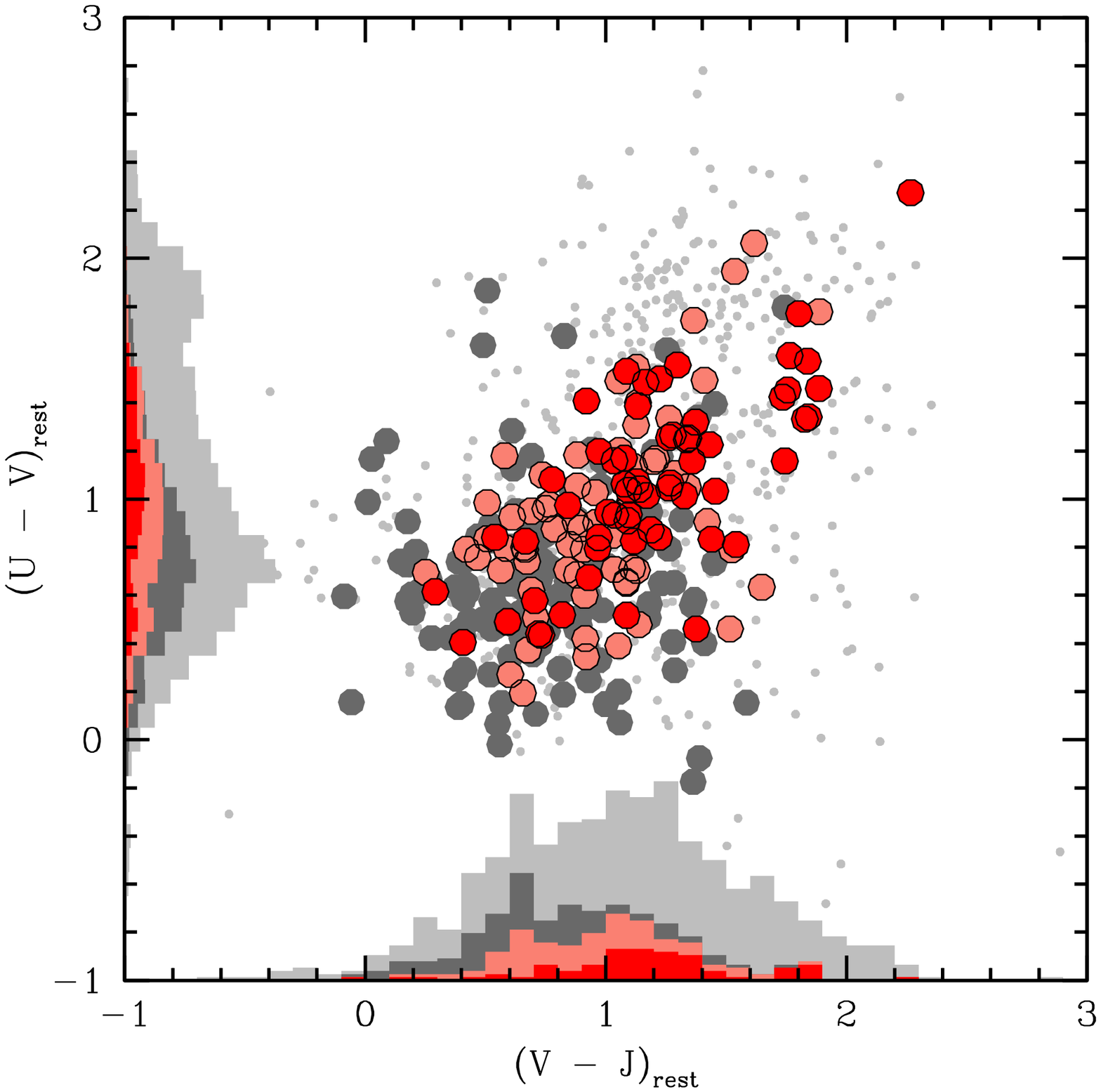}
	\includegraphics[scale=0.43]{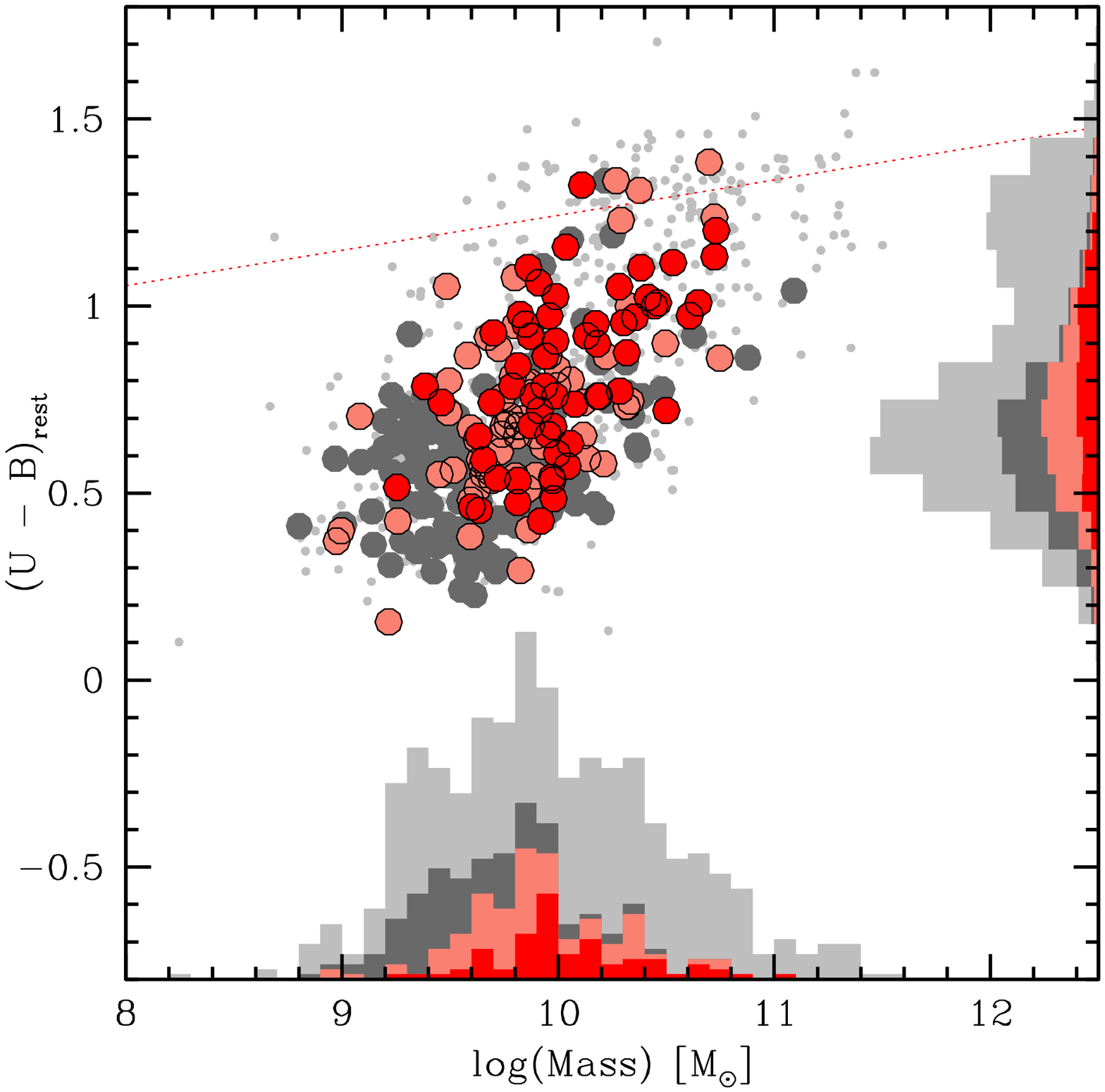}
	\caption{From top-left, clockwise: a) rest-frame (NUV-r)vs.(r-K) colour diagram \citep{arnouts2013};
					  b) rest-frame (U-V)vs.(V-J) colour diagram \citep{williams2009}; 
					  c) rest-frame colour (U-B) vs. stellar mass diagram. The red dotted line is the \emph{red sequence} at $z\sim2$ defined in \citet{cassata2008}.	
		In all plots, small light grey dots represent the GMASS parent sample at $z{\geq}1$, while large points represent the final sample of 286 SFGs analysed in the paper. In particular: dark grey points indicate galaxies without IR data; salmon points indicate galaxies with a 24$\mu m$ detection; red points indicate galaxies with at least one PACS detection. 
		The distribution of each quantity is shown by light grey histograms for the GMASS parent sample ($\#758$), while dark grey histograms represent the distribution over the SFGs sample; salmon histograms represent galaxies with a 24$\mu m$ detection; red histograms represent galaxies with at least one PACS detection.}
	\label{bias}
	\end{figure*}
It is important to establish which part of the galaxy population at $1{<}z{<}3$ is represented by our sample. 
The main bias of our sample comes from two elements: the requirement of a spectroscopic redshift, and the different detection limits of the various sets of photometric data, in particular the IR ones from Spitzer and Herschel. 
To place our galaxies into context, we compared their distributions of mass and colours to those of the full GMASS sample in the same redshift range (Fig. \ref{bias}). Here we note that GMASS is basically a mass-selected sample, whose limiting mass sensitivity over the entire redshift range of our choice ($1{<}z{<}3$) is $log(M/M_{\odot}) {\sim} 10.5$ \citep{kurk2012}. 

Compared to the reference sample, the SFGs sample is probing intermediate-mass objects, the median mass being $M_{\star}{=}10^{9.8} M_{\odot}$, therefore our sample is not mass-complete.
The sample is biased towards bluer and less massive objects, as can be seen from the colour-mass diagram (Fig. \ref{bias} c)), with the reddest and more massive tail populated mainly by the IR-detected galaxies. Objects in the \emph{red sequence} may be either quiescent/passive galaxies or dusty SFGs. The degeneracy can be broken by using colour-colour diagrams (Fig. \ref{bias} a), b)). 
The position of our selection of SFGs in the \emph{(NUV-r)vs.(r-K)} and the \emph{(U-V)vs.(V-J)} diagrams indicates that we have selected galaxies with the highest sSFR, with respect to the parent sample \citep{williams2009, arnouts2013}, of which the IR-detected are the dustier ones.

The different detection limits of the used datasets translate into lower limits in the SFRs that can be probed by a specific indicator. In particular, UV- and optical-based SFR estimators are the deepest ones and allow values down to a few $M_{\odot}yr^{-1}$ to be reached at $z{\sim}1$. Mid-IR data have slightly higher SFR thresholds, while far-IR data are available only for the most star-forming galaxies: $\sim100 M_{\odot}yr^{-1}$ at $z{\sim}2$ and $\sim10 M_{\odot}yr^{-1}$ at $z{\sim}1$ \citep{elbaz2011, rodighiero2011}. 

In the regime of star formation probed by mid- and far-IR data we will derive new relations to correct UV- and optical-based estimators for dust extinction. Then, we will test the robustness of extending such calibrations in lower star formation regimes by comparing one another the newly calibrated estimators for galaxies without IR data. 

Though our sample is inevitably not complete, Fig. \ref{bias} shows that it is fairly representative of a specific galaxy population whose properties may be summarized as follows:
\begin{itemize}
\item redshift between $z\sim1$ and $z\sim3$;
\item intermediate stellar mass (approximately in the range $10^{9.2}{<}M/M_{\odot}{<}10^{10.2}$);
\item blue rest-frame colours: $(NUV-r){<}3 \ \& \  (r-K){<}1.4$ or \\ $(U-V){<}1.2 \ \& \ (V-J){<}1.5$.
\end{itemize}
All the relations derived in the paper should be used within these boundaries.
We also note that the use of our relations is not recommended for galaxies whose spectral continuum indices show strong evidence of old stellar populations ($MgUV{>}1.2$, $C(29-33){>}0.6$, $D4000{>}1.6$) because they could cause an over-correction of dust extinction of either UV or [OII]$\lambda$3727 luminosity and consequently an over-estimate of the total SFR of the galaxy.
	\begin{table}[h!]
	\caption{Number of spectra collected from the different surveys, for the final selection of 286 SFGs. The observed spectral resolution ($R = \lambda/\Delta\lambda$) for each survey is also indicated.}
	\label{spec_survey}
	\centering                          
	\begin{tabular}{l c c c c c}        
	\hline\hline                 
				   &  GMASS  	 & ESO$_{FORS}$ & ESO$_{VIMOS}$ & VVDS 	& K20\\
		   \emph{R}        & 600     	 & 660          & 180,580       & 230  	& 260,380,660\\
	\hline
		            N.	   & 131     	 & 78	      	& 38	       	& 9   	& 30 \\
	\hline\hline
	\end{tabular}
	\end{table}

\section{The benchmark: SFR based on IR luminosity}\label{sec:SFR from IR luminosity}
	\begin{figure}[h!]
	\centering
	\includegraphics[scale=0.43]{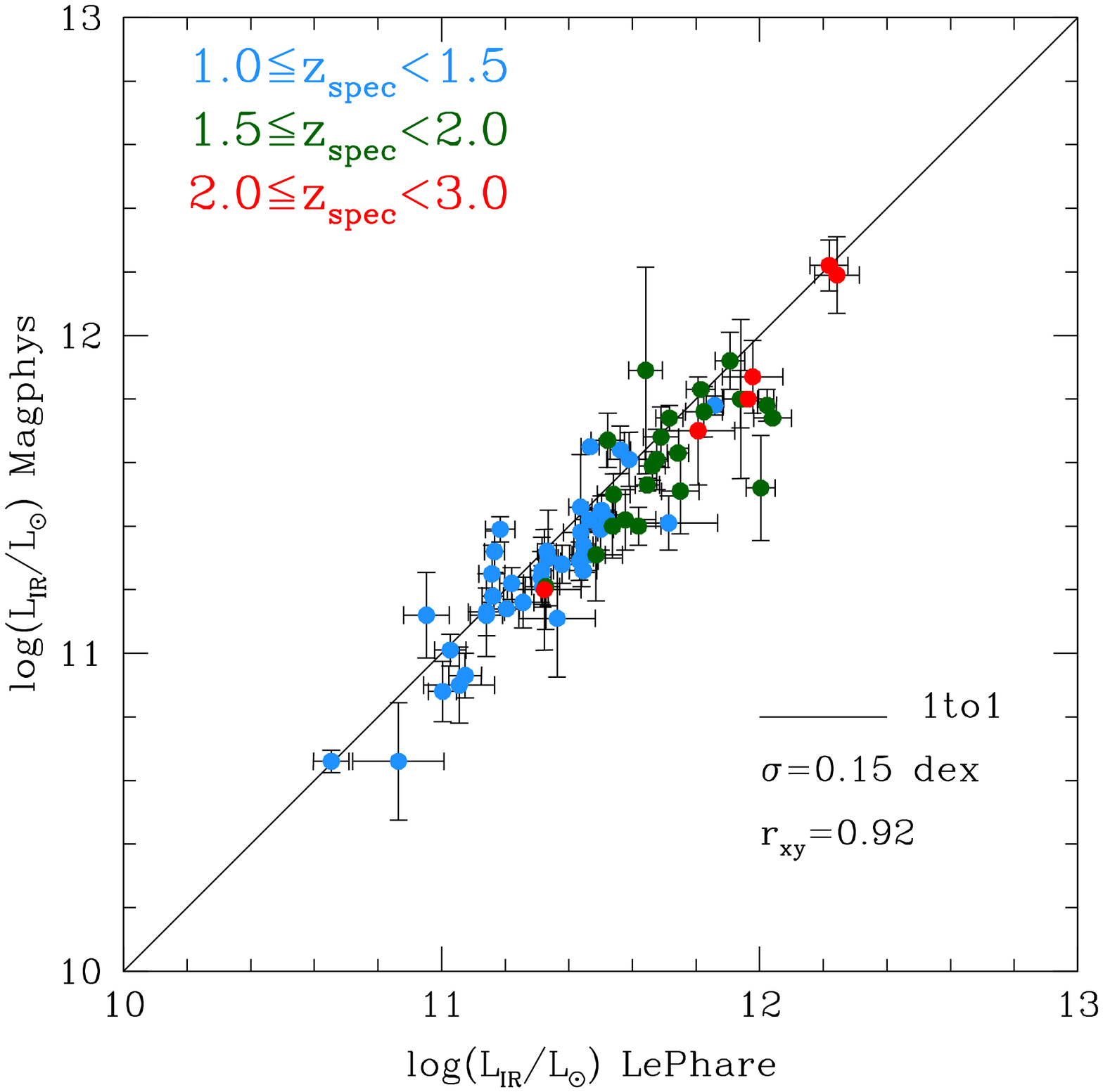}
	\caption{$L_{IR}$ computed using UV to far-IR photometry and the MAGPHYS code vs. $L_{IR}$ computed using Mid- to far-IR photometry and the LePhare code. Points are colour-coded with respect to their redshift. Correlation coefficient $r_{xy}$ (logarithmic scale) is also indicated.}
	\label{LIRvsLIR}
	\end{figure}
The total infrared luminosity ($L_{IR}$) is defined as the integrated luminosity between 8-1000 $\mu m$. To derive $L_{IR}$ for the sources with at least two IR photometric points (i.e. 24 $\mu m$ plus at least one Herschel band), we performed an SED fitting procedure, based on $\chi^{2}$ minimization, using the \emph{MAGPHYS} code in its default configuration \citep{dacunha2008} and all the available photometric information, from U band to PACS data. The inclusion of the whole SED in the fitting procedure allowed us to fully exploit the photometric information. Following an established procedure \citep{rodighiero2010, rodighiero2010b, gruppioni2010, cava2010, popesso2012}, we anchored the spectral fit to the FIR Spitzer-Herschel data-points by increasing the photometric errors of the bands from U to 8$\mu$m (up to 10$\%$)\footnote[4]{For 8 PACS-detected galaxies we were unable to correctly reproduce the observed IR SED. This could indicate the presence of an obscured AGN, whose contribution is unaccounted by our chosen IR SED models. To be conservative, we decided to exclude these galaxies from all further analysis.}. For each source, the 50th percentile of the $L_{IR}$ distribution was taken as the best-value; the quoted uncertainty is the 68th percentile range (${\sim}1{\sigma}$).

To assess the robustness of the resulting quantity, $L_{IR}$ was also computed by fitting the mid- to far-IR photometry alone to a compilation of IR SED template libraries \citep{charyelbaz2001, dalehelou2002, lagache2004, rieke2009} using the \emph{LePhare} code \citep{arnouts1999, ilbert2006}. The two estimates were found to be highly consistent (see Fig. \ref{LIRvsLIR}). This test was in line with the results by \citet{berta2013} who demonstrated that the different methods commonly used in the literature to recover $L_{IR}$ all give consistent estimates when Herschel photometry is implemented in SED fitting. 
	\begin{figure}[b!]
	\centering
	\includegraphics[scale=0.42]{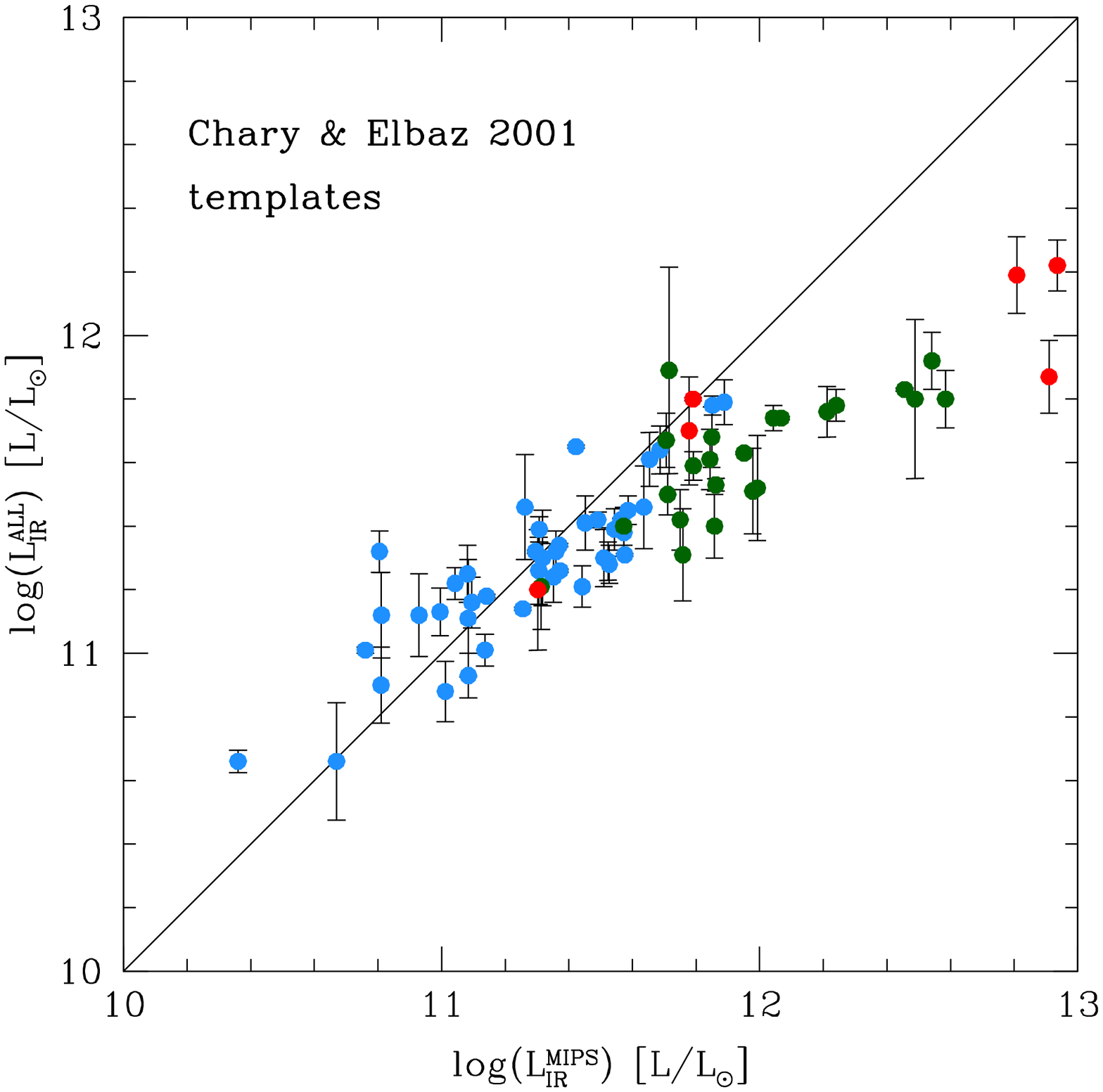}
	\includegraphics[scale=0.42]{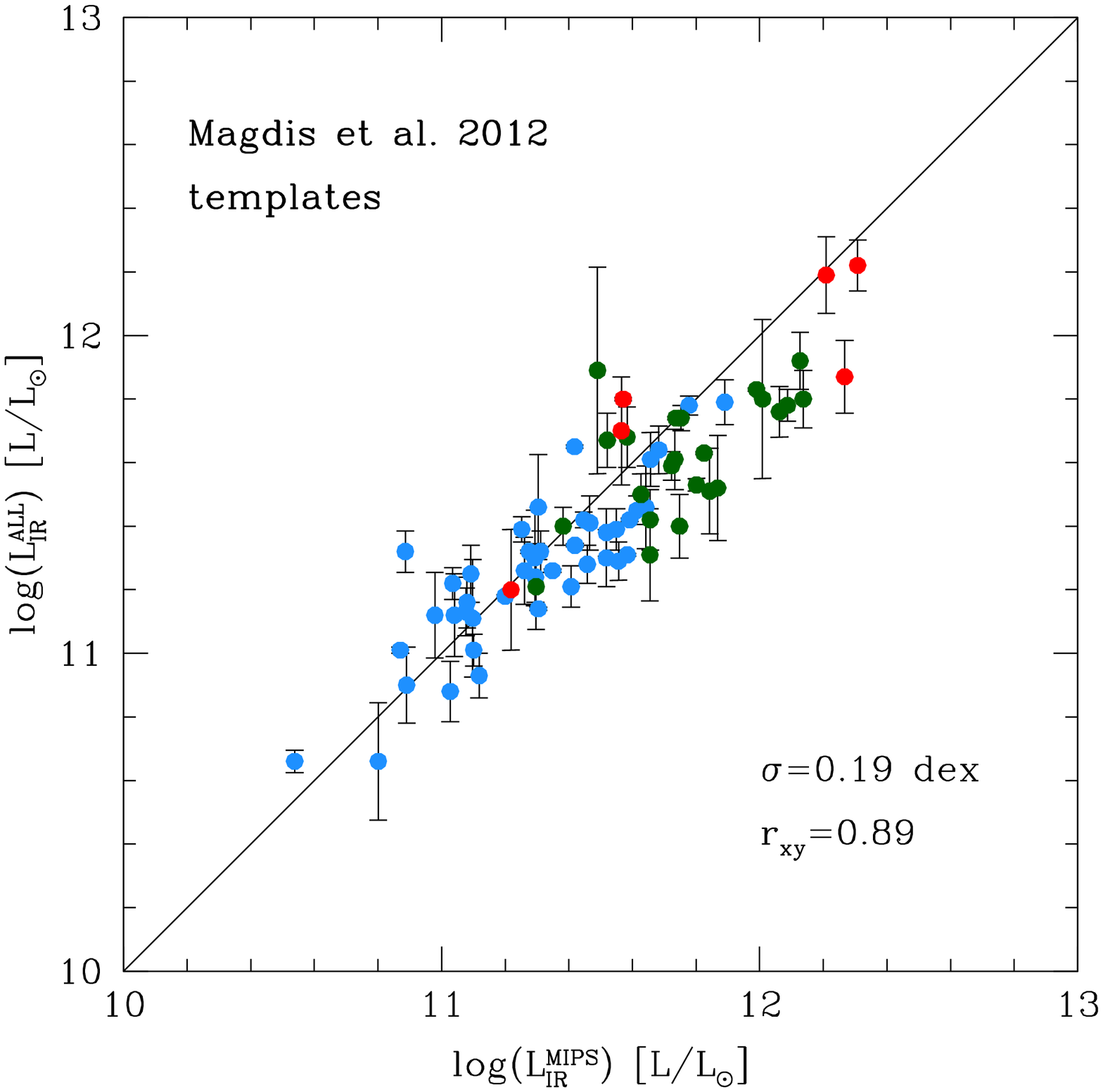}
	\caption{Comparison between $L_{IR}$ computed with \emph{MAGPHYS} using all the available photometric bands (from U-band to PACS), and $L_{IR}$ computed using only the 24 $\mu m$ band and, respectively, \citet{charyelbaz2001} templates (top) and \citet{magdis2012} templates (bottom). Points are colour-coded with respect to their redshift (see Fig. \ref{LIRvsLIR} for the legenda).}
	\label{ALLvs24}
	\end{figure}

To improve the statistics of the sample, especially at low luminosities, $L_{IR}$ was computed also for those galaxies with a detection at 24 $\mu m$ but no PACS data. 
At high redshift ($z {\sim} 2$), locally calibrated IR templates have difficulties in recovering bolometric $L_{IR}$ of highly luminous galaxies from the 24 $\mu m$ flux alone \citep{elbaz2011, nordon2010}. 
Several methods have been proposed to re-calibrate the 24 $\mu m$ flux density \citep{murphy2011, elbaz2011, nordon2013} that make use of the traditional IR templates whose resulting $L_{IR}$ values need to be corrected using various prescriptions. The majority of these conversions from 24 $\mu m$ to $L_{IR}$ either produce trends and biases with the ”true” $L_{IR}$ (defined as the one derived from HERSCHEL data), or include non-linear scaling that strongly increase the scatter \citep{nordon2013}.

A different approach is to build brand new SED templates taking advantage of the most recent far-IR and sub-mm data. This latter method is preferable since it does not need to apply a posteriori corrections to the resulting $L_{IR}$ values. Here we adopted the main sequence templates of \citet{magdis2012} to extrapolate $L_{IR}$ from the 24 $\mu m$ flux densities\footnote[5]{We did not use the \emph{MAGPHYS} code to derive $L_{IR}$ for galaxies with only a 24 $\mu m$ detection (and no PACS data) because the code requires IR data at longer wavelengths to reliably estimate $L_{IR}$.}. We tested this choice for the sub-sample of PACS-detected galaxies trough the comparison between $L_{IR}$ derived from 24 $\mu m$ alone and $L_{IR}$ computed using all available photometric data from U-band to PACS, as explained in the previous paragraph. The result of this test is shown in Fig. \ref{ALLvs24} where we also show, for comparison, the output of using the popular \citet{charyelbaz2001} templates. The \citet{charyelbaz2001} templates work well up to $z{\sim}1$, but at higher redshifts they overestimate the true $L_{IR}$, mainly because at such redshifts polycyclic aromatic hydrocarbon (PAH) emissions contribute significantly to the 24 $\mu m$ band flux and are difficult to model \citep{elbaz2011, nordon2010}. With the \citet{magdis2012} templates the resulting 24 $\mu m$-derived $L_{IR}$ is much more consistent with our true estimate of total $L_{IR}$ at all luminosities, though there is still a slight overestimate of the total $L_{IR}$ at $z{\geq}1.5$, that we attribute to the PAH contribution to broad-band flux. We quote the mean scatter from the 1-to-1 relation (0.2 dex) as the error on the estimate of $L_{IR}$ when only the 24 $\mu m$ band is available.

\begin{figure}[h!]
	\centering
	\includegraphics[scale=0.43]{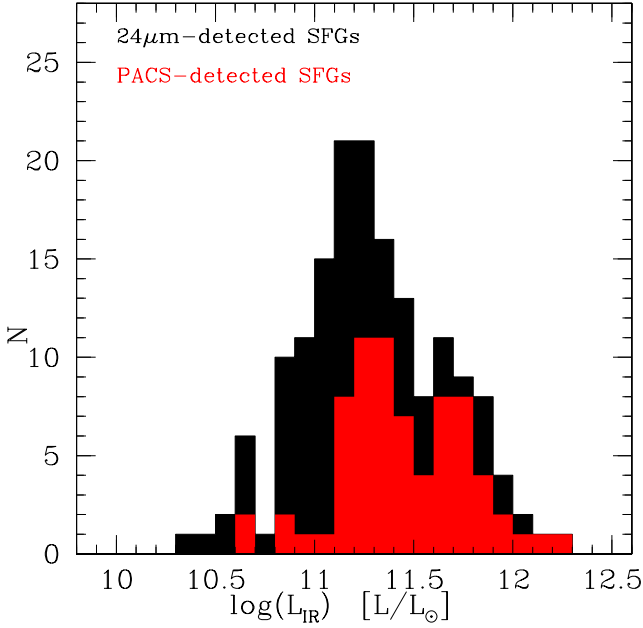}
	\caption{Distribution of $L_{IR}$ over the sample of SFGs with at least a 24 $\mu m$ detection (black histogram). The red histogram shows the $L_{IR}$ distribution of PACS-detected sources only.}
	\label{lir_hist}
	\end{figure}

In the rest of the paper, when naming $L_{IR}$ we will be referring to the $L_{IR}$ estimated from SED fitting to U-to-PACS photometry using the \emph{MAGPHYS} code for galaxies with at least two IR photometric points (i.e. 24 $\mu m$ plus at least one Herschel band), while we will be referring to the $L_{IR}$ estimated from the \citet{magdis2012} templates for galaxies with only a detection at 24 $\mu m$ but no PACS data. Fig. \ref{lir_hist} shows the final $L_{IR}$ distribution for the sample of SFGs. 

The infrared luminosities can be converted into SFR using the \citet{kennicutt1998} relation\footnote[6]{The original relation assumes a Salpeter IMF. We scaled it by a factor of 1.55 to match our chosen IMF \citep{kroupa2001}.}: 
\begin{equation} 
SFR_{IR} (M_{\odot}yr^{-1}) = 2.9 \times 10^{-44} L_{IR} (erg/s),
\end{equation} 
In star-forming galaxies $SFR_{IR}$ is expected to be a good estimate of the total SFR. However, there might be a fraction of UV radiation that escapes dust absorption and that is not incorporated into the calibrations. This missing unattenuated component varies from essentially zero in dusty starburst galaxies to nearly 100$\%$ in dust-poor dwarf galaxies and metal-poor regions of more-massive galaxies \citep{kennicutt2012}. To overcome this bias many authors suggest adding another term to $SFR_{IR}$ to account for the unobscured UV light \citep{papovich2007, rodighiero2010, wuyts2011, nordon2013, kennicutt2012}. This second term is simply the SFR derived from UV luminosity\footnote[7]{$L_{\nu}(1500\AA)$ of galaxies in the \emph{UV sample} was measured directly on rest-frame spectra, corrected for slit-losses. In the \emph{[OII] sample} the range covered by spectra is shifted towards longer wavelengths: in these cases, $L_{\nu}(1500\AA)$ was measured by computing the absolute magnitude in a squared filter X (300$\AA$ wide, centred at 1500$\AA$), using the observed magnitude in the filter Y, which is chosen to be the closest to $\lambda(X)\times(1+z)$.}, not corrected for dust extinction. We adopted the appropriate relation from \citet{kennicutt1998},
\begin{equation}
SFR_{UV} (M_{\odot}yr^{-1}) = 0.9 \times 10^{-28} L_{\nu}(1500\AA) (erg s^{-1} Hz^{-1}),
\end{equation}
and finally define 
\begin{equation}
SFR_{IR+UV}{=}SFR_{IR}+SFR_{UV}
\end{equation}
as total SFR (obscured plus un-obscured) \citep{wuyts2011, nordon2013}. 

\begin{figure}[b!]
	\centering
	\includegraphics[scale=0.48, trim=20mm 0mm 20mm 0mm, clip=true]{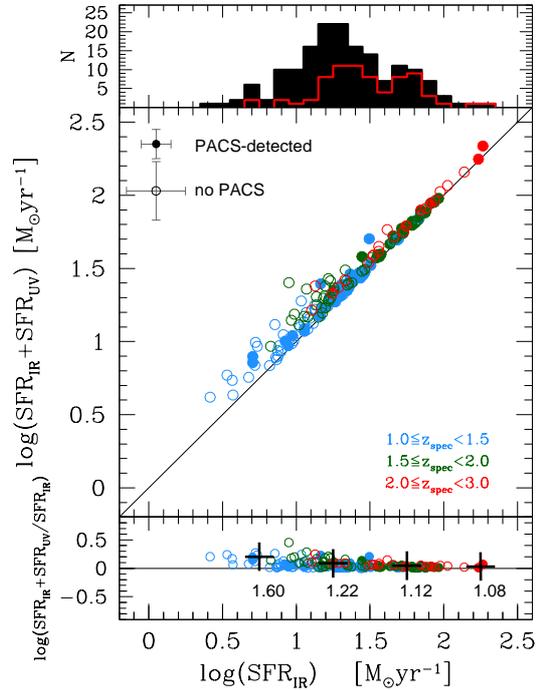}
	\caption{Main plot: comparison between $SFR_{IR+UV}$ and $SFR_{IR}$ for SFGs with at least one IR detection. Filled points mark PACS-detected galaxies. Bottom plot: $SFR_{IR+UV}$ over $SFR_{IR}$ ratio as a function of $SFR_{IR}$. Median ratio in bins of $SFR_{IR}$ are also shown. Upper plot: black histogram is the distribution of $SFR_{IR}$ over the entire sample of SFGs; red empty histogram is the distribution of PACS-detected sources only.}
	\label{sfriruvVSsfrir}
	\end{figure}

In Fig. \ref{sfriruvVSsfrir} we plot the total SFR ($SFR_{IR+UV}$) vs. the IR component alone. 
The contribution of $SFR_{UV}$ to the total SFR is almost negligible in ultra-luminous (ULIRGs: $L_{IR}{\slash}L_{\odot}{\geq}10^{12}$) and the majority of luminous (LIRGs: $10^{11}{<}L_{IR}{\slash}L_{\odot}{<}10^{12}$) IR galaxies, while in normal SFGs ($L_{IR}{\slash}L_{\odot}{\leq}10^{11}$) its contribution to the total SFR grows to about $30\%$. However, we have tested how neglecting this contribution would alter our results and we found that none of the calibrations derived in this paper would significantly change if $SFR_{IR}$ were assumed as total SFR, instead of $SFR_{IR+UV}$.

In the following sections we shall compare $SFR_{IR+UV}$ to the other SFR estimators available for our sample, with particular emphasis on spectroscopic data. 
Using the IR data to calibrate UV and optical estimators, we can derive a set of relations that can be used to consistently compare galaxies in a wide range of redshifts ($1{<}z{<}3$) and with different kinds of available data (from observed optical photometry and spectra, to IR data). 

\section{Dust-extinction correction: SFR from UV continuum luminosity}\label{sec:SFR from UV flux}

The UV luminosity emitted by young stars is a direct indicator of ongoing star formation. However, it is typically severely extincted by the dust surrounding star-forming regions, therefore the measured rest-frame UV luminosity must be corrected before being converted into SFR. Empirical relations are usually applied for this purpose, most of them calibrated on local galaxies. 

\subsection{UV sample: SFGs at $1.6{<}z{<}3$}\label{sec:The UV continuum slope}

In galaxies dominated by a young stellar population, the shape of the UV continuum can be fairly accurately approximated by a power law $F_\lambda {\varpropto} \lambda^\beta$, where $F_\lambda$ is the observed flux ($erg~s^{-1}~cm^{-2}~\AA^{-1}$) and $\beta$ is the continuum slope \citep{calzetti1994}. \citet{meurer1999} found, for a sample of local starburst galaxies, a correlation between the UV spectral slope ($\beta$) and the ratio of $L_{IR}$ over UV luminosity ($L_{UV}=\nu L_{\nu}(1500\AA)$), not corrected for extinction, which can be translated into attenuation $A_{1600~\AA}$. \citet{overzier2011} and \citet{takeuchi2012} independently revised the original \citet{meurer1999} relation correcting for the effect of the small \emph{IUE} aperture used in the original calibration, thanks to \emph{GALEX} data, and using an update estimate for $L_{IR}$ (total IR emission instead of \emph{IRAS} far-IR (FIR) emission in the range $40 - 120 \mu m$). The attenuation vs. $\beta$ relation has since been used for galaxies at various redshifts, though different studies have reported contradicting results about its general validity \citep{calzetti2001, boissier2007, seibert2007, reddy2010, overzier2011, reddy2012, takeuchi2012, buat2012, heinis2013, nordon2013, castellano2014, oteo2013, oteo2014}. Here we tested the relation for our sample using $L_{IR}$ derived from our mid- and far-IR photometry, in order to obtain an estimate of SFR from dust-corrected UV flux consistent with the one from IR+UV defined in the previous section.

The attenuation is defined from the ratio between $L_{IR}$ and uncorrected $L_{UV}$ ($IRX=L_{IR}/L_{UV}$). However, the exact definition is not unique across the literature \citep{meurer1999, buat2005}. Here we adopt the definition \citep{nordon2013}
\begin{equation} log(SFR_{IR+UV})=log(SFR_{UV})+0.4~A_{IRX}
\end{equation}
\begin{equation}
A_{IRX}~=~2.5~log\left(\frac{SFR_{IR}}{SFR_{UV}}~+~1\right),
\end{equation}
where $SFR_{UV}$ is the uncorrected quantity, as defined in Sec. \ref{sec:SFR from IR luminosity}, and $SFR_{IR+UV}$ is the total SFR. This definition is consistent with the ones adopted by \citet{overzier2011} and \citet{takeuchi2012}.

The slope $\beta$ is defined in the wavelength range from ${\sim}1250{\AA}$ to ${\sim}2600{\AA}$ \citep{calzetti1994}, and can be measured either directly on spectra or by using photometry. In our sample, only galaxies in the \emph{UV sample} have spectra with the right wavelength coverage; for the \emph{[OII] sample} photometry has to be used. 

\begin{figure}[h!]
	\centering
	\includegraphics[scale=0.43, trim=0mm 0mm 0mm 40mm, clip=true]{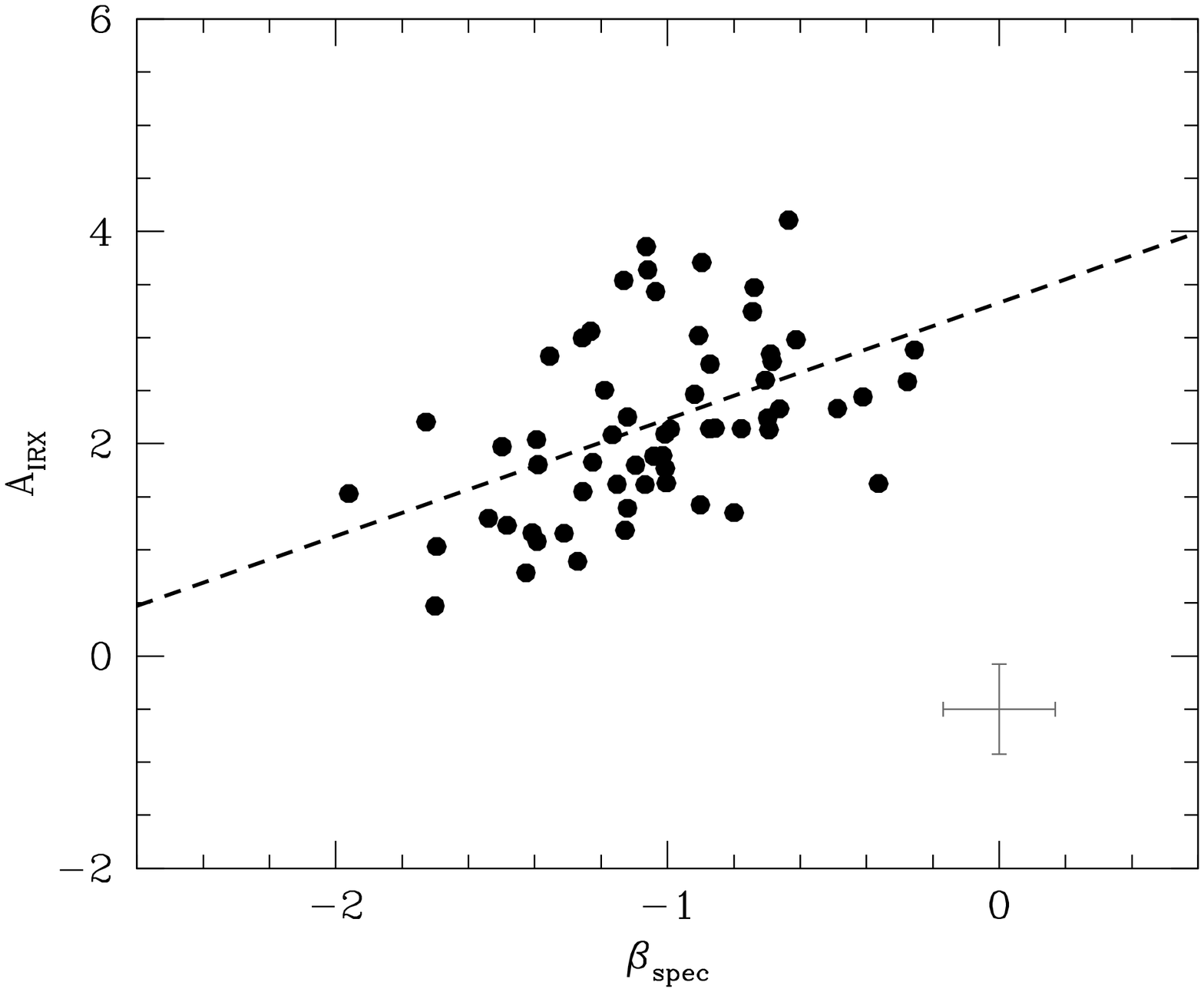}
	\includegraphics[scale=0.43, trim=0mm 0mm 0mm 40mm, clip=true]{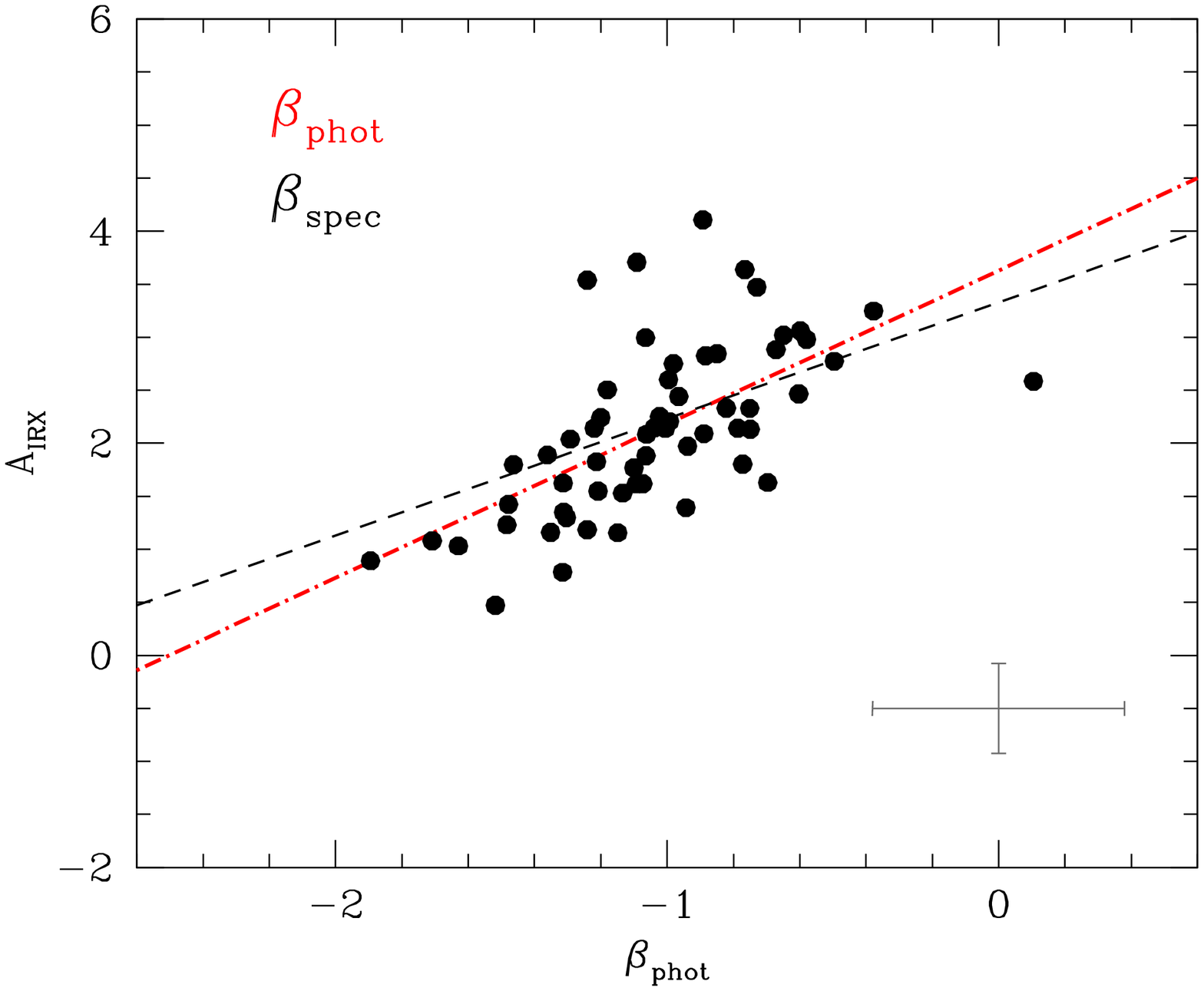}
	\caption{$A_{IRX}$ vs. $\beta$ for IR-detected galaxies in the \emph{UV sample}. Top: $\beta$ is measured from the spectra. The black line is a fit to the data (Eq.6). Bottom: $\beta$ is measured from photometry. The red line is a fit to the data (Eq.7). The black line is the relation for $\beta_{spec}$.}
	\label{slope_vs_irx}
	\end{figure}

The 62 IR-detected galaxies with a measurable spectral slope were then used to test the $A_{IRX}$ vs. $\beta$ relation at $1.6{<}z{<}3$ and the result is shown in Fig. \ref{slope_vs_irx}. The continuum slope of each spectrum was derived through a linear fit of the average fluxes in the spectral windows defined by \citet{calzetti1994}, in the $log(F_\lambda)-log(\lambda)$ plane. More details about the spectroscopic determination of $\beta$ can be found in the Appendix.  

We then derive the best-fit relation by performing a linear fit of the form $A_{IRX} = C_{0} \times \beta + C_{1}$ to the data,
\begin{equation}
A_{IRX}~=~(1.10 \pm 0.23) \times \beta_{spec} + (3.33 \pm 0.24)
\end{equation}
with an r.m.s. dispersion of the residuals ${\sim}0.7$. 
 
Our relation ($C_{0}{=}1.10$) is broadly consistent with the one by \citet{murphy2011} for a sample of $24 \mu$m-detected galaxies at $z{\sim}2$ ($C_{0}{=}0.75$). On the other hand, it is flatter than in the UV-selected samples of \citet{overzier2011} (local) and \citet{buat2012} ($z{\sim}1.5$) (respectively, $C_{0}{=}1.96$ and $C_{0}{=}1.70$). Also, the standard deviation in the latter cases is of the order of $\sim$0.3 dex, while our relation is more dispersed. We ascribe it mainly to the different sample selections. A low dispersion is expected when galaxies are UV-selected local starbursts or high-redshift equivalents, but deviations from a tight relation have been reported both for dusty IR-luminous galaxies and for quiescently star-forming galaxies \citep[and references therein]{buat2012, takeuchi2012}. Our parent catalogue is mass-selected, therefore our sample is expected to include objects with a wide range of star formation histories and different levels of starburst activity that may be characterized by different attenuation laws and increase the scatter in the relation, with respect to a UV selection. In particular, we found that the objects that deviate the most from our relation (i.e. with the highest $A_{IRX}$ at fixed $\beta$) have the highest $L_{IR}$. 

The continuum slope can be derived also from photometric data, either by measuring $\beta$ directly from the best-fit SED template to the overall photometric data of the galaxy \citep[e.g.][]{oteo2013, oteo2014}, or by fitting the photometric points which sample the UV continuum at the redshift of the galaxy with a power-law function \citep[e.g.][]{bouwens2009, buat2012, nordon2013, pannella2014}. 
We decided to use the latter method, to avoid model dependencies. Uncertainties were computed by randomizing the observed fluxes according to their errors and repeating the fit to derive a new slope.
Since even with the photometry we cannot probe wavelengths as short as ${\sim}1200\AA$ at $z{\sim}1$, for the photometric determination of the continuum slope we decided to use, at all redshifts, a shorter rest-frame wavelength range to derive $\beta_{phot}$, starting from $\lambda_{rest}{\sim}1500\AA$. This wavelength coverage is also the most popular in the literature, when the slope is derived from photometry and not from spectra \citep[see e.g.][]{bouwens2009, bouwens2012, nordon2013, castellano2012}. 

We computed $\beta_{phot}$ in the \emph{UV sample} for the same galaxies for which $\beta_{spec}$ had been already derived and a similar relation was found between $\beta_{phot}$ and $A_{IRX}$,
\begin{equation}
A_{IRX}~=~(1.45 \pm 0.25) \times \beta_{phot} + (3.63 \pm 0.26),
\end{equation}
with an r.m.s. dispersion of the residuals ${\sim}0.6$. 
This relation gives attenuation values that are consistent with those derived from spectroscopy. This result will justify us to use the photometric slope to derive a similar relation for the galaxies in the \emph{[OII] sample}, where the spectra do not cover the wavelength range needed to compute $\beta$.

\subsection{A little digression about regression methods}\label{sec:A little digression}
Since our aim is to provide a relation to predict the value of a variable from the measurement of another, we adopted an OLS (Y$|$X) fit in the derivation of the $A_{IRX}$ vs. $\beta$ relation. The OLS (Y$|$X) fit is the most widely used in the literature and we are assuming that "OLS" and "linear regression", in the absence of further specification, means OLS (Y$|$X) in all the works whose calibrations we cite here to compare with our findings. 

A fit that treats symmetrically the two variables is instead to be preferred when the goal is to estimate the underlying functional relation between the variables \citep{isobe1990}. The slope of the $A_{IRX}$ vs. $\beta$ relation ($C_{0}$) depends on the attenuation curve. The $C_{0}$ parameter expected from a given attenuation curve should be compared to the value derived from a symmetric fit, instead of an OLS (Y$|$X). 
We report that an OLS \emph{bisector} fit \citep{isobe1990} to the data analysed in Fig. \ref{slope_vs_irx} (top) would produce the following relation (see Eq. 6 for the corresponding OLS (Y$|$X) fit)
\begin{equation}
A_{IRX}~=~(1.96 \pm 0.23) \times \beta_{spec} + (4.22 \pm 0.19) \ \ \ \ \ [OLS Bisector],
\end{equation}
while in the case of $\beta_{phot}$ (Fig. \ref{slope_vs_irx} bottom) we would have
\begin{equation}
A_{IRX}~=~(2.16 \pm 0.27) \times \beta_{phot} + (4.36 \pm 0.27) \ \ \ \ \ \ [OLS Bisector],
\end{equation}
whose corresponding OLS (Y$|$X) fit is given in Eq. 7.
For comparison, the $C_{0}$ value expected from the Calzetti law is $C_{0}{=}2.3$, which gives a relation fairly consistent with our findings.

\subsection{Extinction from observed colour}\label{sec:Extinction from observed colour}

Another popular empirical recipe to derive a dust extinction correction to be applied to UV fluxes, for galaxies at $z{\geq}1.4$, comes from the tight correlation that \citet{daddi2004} found between the observed $(B-z)$ colour and $E(B{-}V)$ of \emph{BzK} galaxies. This relation is valid only for galaxies in the redshift range $1.4{<}z{<}2.5$, where the $(B-z)$ colour is basically another way of expressing the UV continuum slope. In the \citet{daddi2004} paper, the $E(B{-}V)$ is derived from SED fitting to UV-to-NIR (near-IR) multi-band photometry (assuming \citet{bruzual2003} models, Salpeter IMF, constant star formation, and a Calzetti law).
	\begin{figure}[h!]
	\centering
	\includegraphics[scale=0.43]{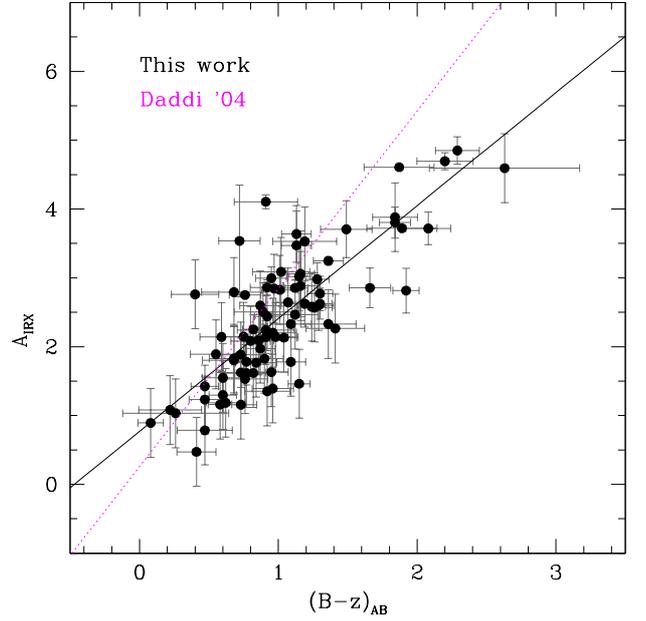}
	\caption{$A_{IRX}$ vs. the (B-z) colour. Black line is a fit to the data. The magenta line is the \citet{daddi2004} relation, plotted as reference.}
	\label{bz_vs_airx}
	\end{figure}

In this work we use instead the available IR data to calibrate the $(B-z)$ colour as an extinction estimator. A linear fit to the data produces the following relation:
\begin{equation}
A_{IRX}~=~(1.64 \pm 0.13) \times (B-z)_{AB} + (0.77 \pm 0.13).
\end{equation}
In Fig. \ref{bz_vs_airx} we plot all IR-detected SFGs at $1.4{<}z{<}2.5$. The original \citet{daddi2004} relation is shown as a reference\footnote[8]{The \citet{daddi2004} relation is E(B-V) vs. (B-z). In the plot, E(B-V) was converted to $A_{IRX}$ assuming a Calzetti law, as in the original calibration: $E(B{-}V){=}A_{IRX}/k_{1500}$, where $k_{1500}$ is the value of the reddening curve at 1500$\AA$ \citep{calzetti2000}.}. The two relations are in general quite consistent, but ours is flatter for galaxies with reddest \emph{(B-z)}. 

The attenuation derived from $(B-z)$ colour is consistent with the one derived from the continuum slope in the previous section, not only for the IR-detected galaxies (that were actually used to derive both sets of calibrations), but also for galaxies without IR information. 

\subsection{[OII] sample: SFGs at $1{<}z{<}1.6$}\label{sec:The OII sample}
	\begin{figure}[h!]
	\centering
	\includegraphics[scale=0.43, trim=0mm 0mm 0mm 40mm, clip=true]{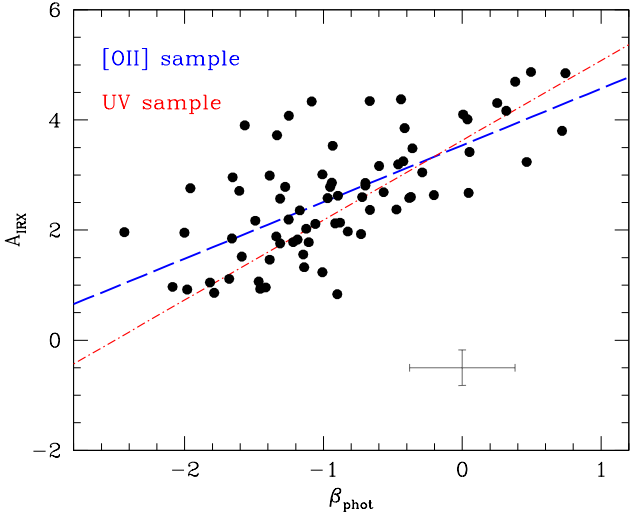}
	\caption{$A_{IRX}$ vs. $\beta_{phot}$ for IR-detected galaxies in the \emph{[OII] sample}. The blue line is a fit to the data (Eq. 11). The red line is the $A_{IRX}$ vs. $\beta_{phot}$ relation for the \emph{UV sample} (Eq. 7).}
	\label{photslope_vs_airx_oii}
	\end{figure}
	\begin{figure}[h!]
	\centering
	\includegraphics[scale=0.53, trim=20mm 20mm 0mm 0mm, clip=true]{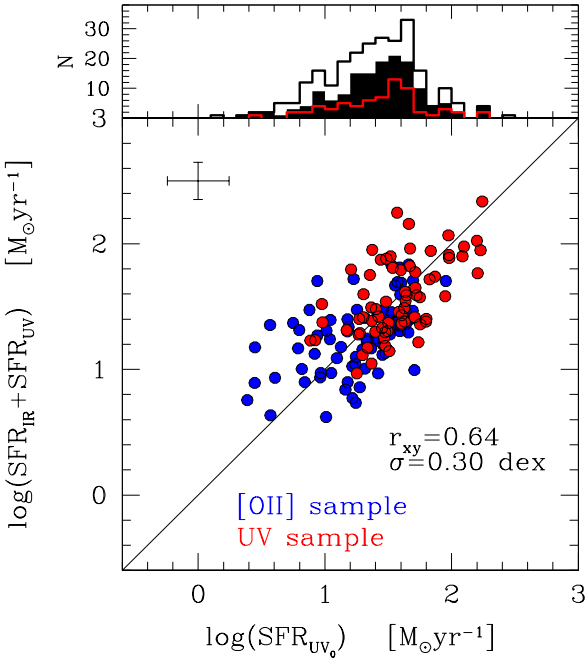}
	\caption{Main plot: comparison between $SFR_{IR+UV}$ and $SFR_{UV_{0}}$ corrected using the $A_{IRX}$ vs. $\beta$ relations derived in the paper. $\beta$ is derived from spectroscopy, when possible, otherwise from photometry (see the text for more details). Upper plot: black empty histogram is the distribution of $SFR_{UV_{0}}$ over the entire sample of SFGs; black filled histogram is the distribution of $SFR_{UV_{0}}$ over the sample of 24 $\mu m$-detected sources (both with and without PACS data); red empty histogram is the distribution of PACS-detected sources only.}
	\label{sfriruvVSsfruv_slope}
	\end{figure}

In our sample, only galaxies in the \emph{UV sample} have spectra with the right wavelength coverage to compute the continuum slope. However, photometry can also be used for the same task. For the galaxies in the \emph{[OII] sample} we computed the slope $\beta_{phot}$ between $\sim$1500$\AA$ and $\sim$2600$\AA$, as explained in the previous section, and we found the following relation with $A_{IRX}$,
\begin{equation}
A_{IRX}~=~(1.03 \pm 0.26) \times \beta_{phot} + (3.54 \pm 0.25),
\end{equation}
with an r.m.s. dispersion of the residuals is ${\sim}0.8$ (Fig. \ref{photslope_vs_airx_oii})\footnote[9]{The OLS \emph{bisector} fit gives $C_{0}{=}1.49$}. Comparing Eq. 7 and Eq. 11 we find that in the \emph{[OII] sample} the slope of the relation ($C_{0}$) is slightly flatter than in the \emph{UV sample}. However, the difference is not highly significant, therefore we see no evolution of the $A_{IRX}$ vs. $\beta$ relation between $z{\sim}1.3$ and $z{\sim}2.3$. Moreover, dust attenuation of the galaxies in both samples broadly follows the prediction of the Calzetti law (when considering the OLS \emph{bisector}  relations; see Sec. \ref{sec:A little digression}), though in the low-redshift bin the Calzetti law tends to over-predict the attenuation in objects with the reddest slope, with respect to the results of our calibrations \citep[see also][]{pannella2014}.

The relations derived in this and the previous sections were applied to all the galaxies in our sample, choosing the appropriate one for each object to derive $A_{IRX}$ values from the UV continuum slope and compute the un-extincted SFR ($SFR_{UV_{0}}$). The final check to assess how good the calibrations are to recover the total SFR comes from the comparison between $SFR_{UV_{0}}$ and $SFR_{IR+UV}$ in Fig. \ref{sfriruvVSsfruv_slope} where we can see that the two estimates are in good agreement.

\section{Dust-extinction correction: SFR from [OII]$\lambda$3727 emission line luminosity}\label{sec:SFR from [OII] emission line}

The next SFR estimator to be considered was the forbidden line [OII]$~\lambda 3727$. 
The luminosities of forbidden lines are not directly coupled to the ionizing luminosity, and their calibration as SFR indicators is sensitive to density variations in dust reddening, chemical abundance, and ionization among star-forming galaxies \citep{kennicutt1998, jansen2001, kewley2004, moustakas2006}.
The excitation of [OII]$\lambda$3727 is sufficiently well behaved that it can be calibrated empirically as a quantitative SFR tracer through H$\alpha$, though the [OII]/H$\alpha$ ratio in individual galaxies can vary considerably and this is the main uncertainty related to the [OII]-derived estimate of SFR.
The other two major sources of uncertainty are the effects of metallicity and dust extinction on the conversion of [OII]$\lambda$3727 luminosity into SFR. 

Various approaches have been proposed in the literature to derive dust-corrected estimates of $SFR_{[OII]_{0}}$. \citet{kewley2004}, for example, derive a local $SFR_{[OII]_{0}}$ calibration based on the intrinsic (e.g. reddening corrected) [OII]$\lambda$3727 luminosity and the abundance. The local calibration by \citet{moustakas2006} is instead parametrized in terms of the B-band luminosity to remove the systematic effects of reddening and metallicity. Both calibrations use as reference SFR the estimate from H$\alpha$ luminosity, corrected for dust-extinction using the Balmer decrement.

\subsection{The [OII]$\lambda$3727 equivalent width vs. $\beta$ relation}\label{sec:The equivalent width vs. beta relation}

Here we exploit the Spitzer-Herschel data and use $SFR_{IR+UV}$ as reference to calibrate a dust-corrected estimate of SFR from [OII]$\lambda$3727 luminosity for our \emph{[OII] sample} of galaxies at $1.0{<}z{<}1.6$.

\begin{figure*}[t!]
	\centering
	\includegraphics[scale=0.68, trim=0mm 100mm 0mm 0mm, clip=true]{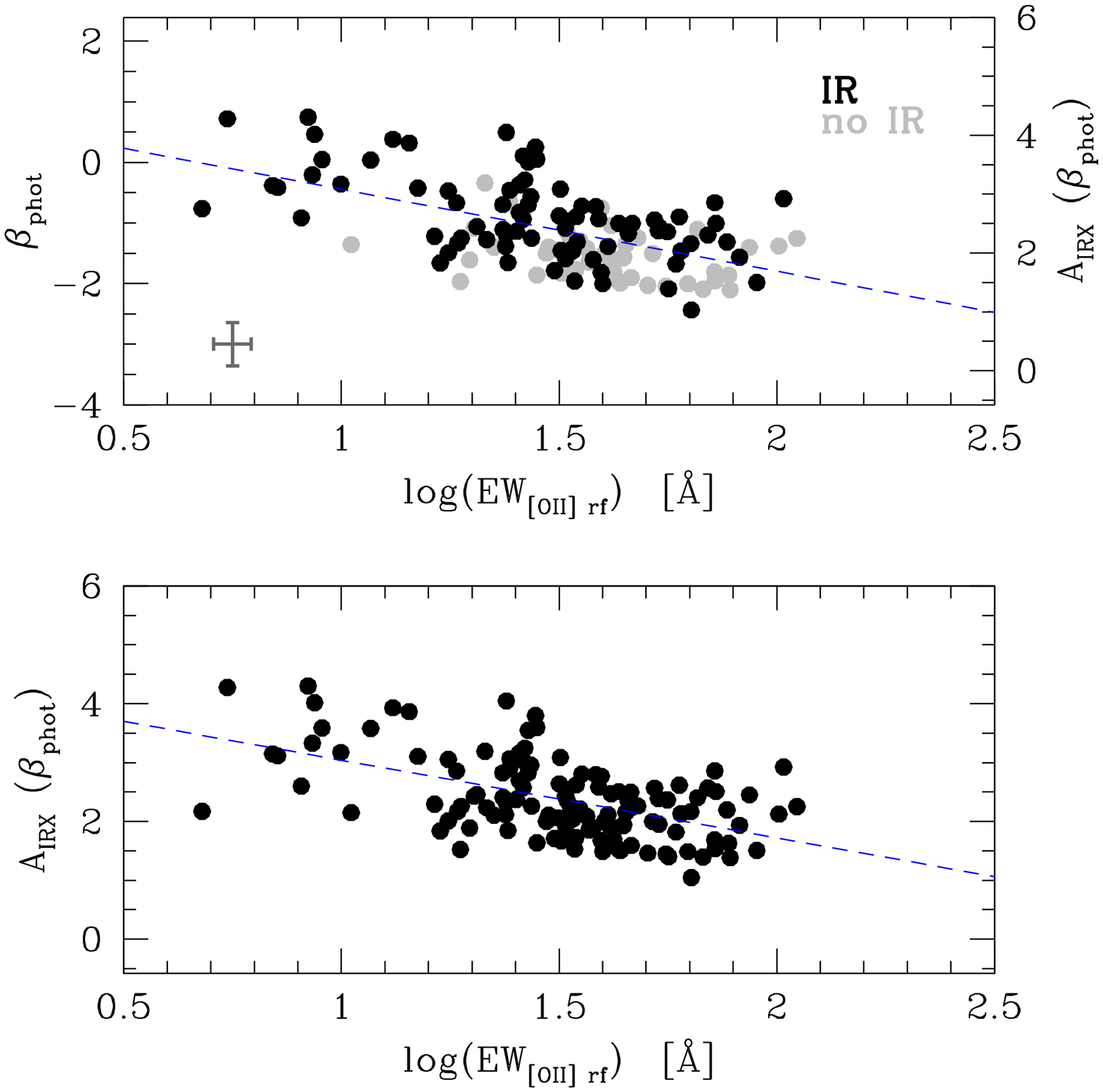}	
	\includegraphics[scale=0.68, trim=0mm 50mm 0mm 10mm, clip=true]{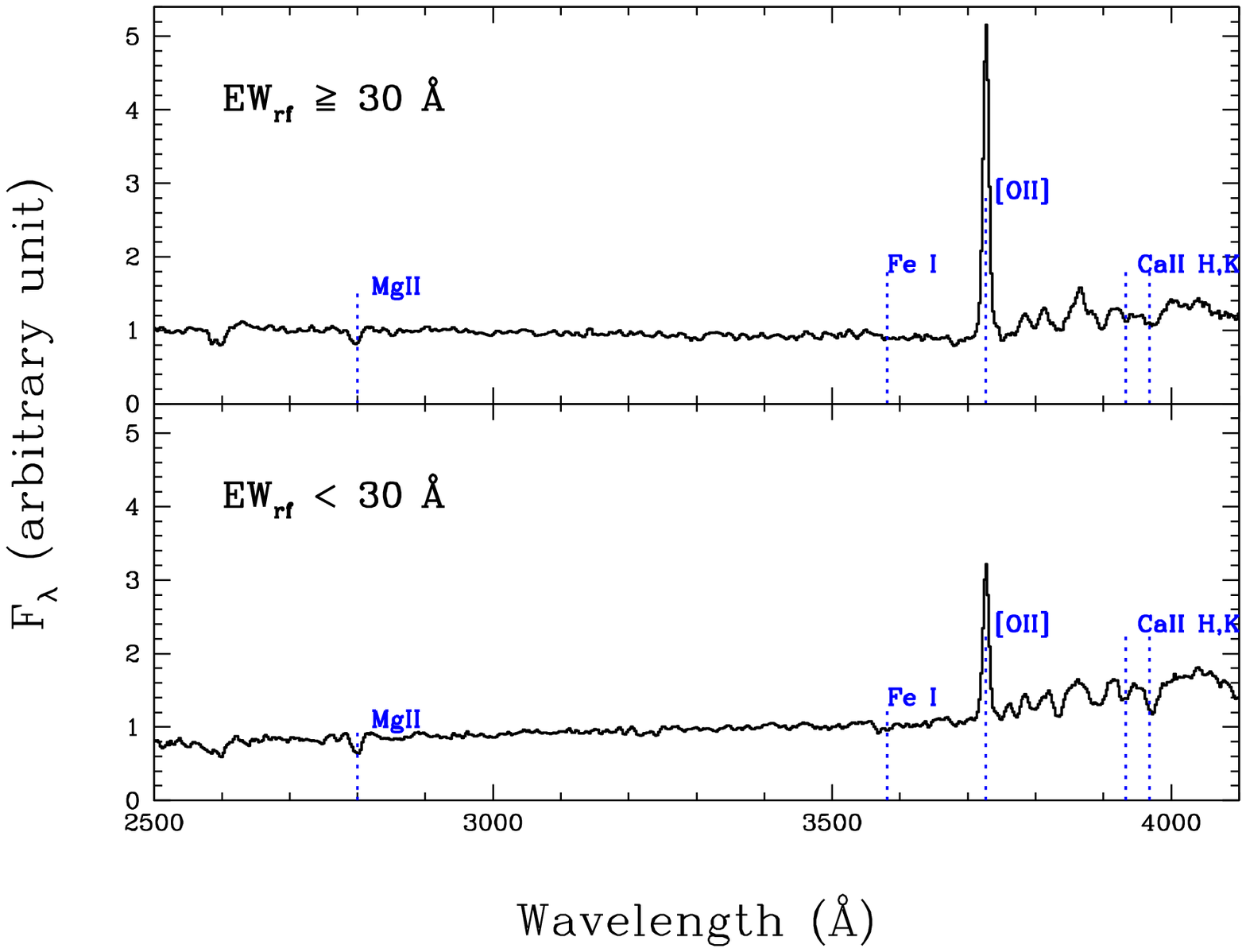}
	\caption{Top: $\beta_{phot}$ vs. log($EW_{[OII]}$) rest-frame. We use the convention of positive EW for emission lines. The right axis converts $\beta_{phot}$ to $A_{IRX}$ using Eq. 11. Black points represent IR-detected galaxies, while grey points represent galaxies with no IR detection. The blue dashed line is a linear fit to the points. Bottom: average spectra in two bins of rest-frame $EW_{[OII]}$. Lines of interest are labelled.} 	
	\label{airx_oii}
	\end{figure*}

\begin{figure}[h!]
	\centering
	\includegraphics[scale=0.33]{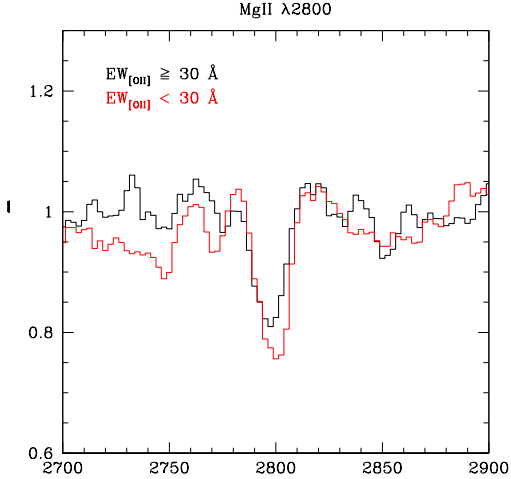}
	\caption{Composite spectra of galaxies in the \emph{[OII] sample} in two bins of rest-frame $EW_{[OII]}$: zoomed-in image of the MgII$\lambda$2800 ISM absorption line.}
	\label{mgii}
	\end{figure}

Integrated line fluxes of [OII]$\lambda$3727 were measured and corrected for slit-losses using the available photometry. We assume the \citet{kennicutt1998} calibration (scaled to a Kroupa IMF) to convert [OII]$\lambda$3727 luminosity into SFR,
\begin{equation}
SFR_{[OII]_{0}} (M_{\odot}yr^{-1}) = 0.9 \times 10^{-41} L_{[OII]_{0}} (erg s^{-1}),
\end{equation}
where $L_{[OII]_{0}}$ is the intrinsic [OII]$\lambda$3727 luminosity, corrected for dust extinction.
We explored the possibility of calibrating a relation between attenuation, derived by our IR data, and some property of the [OII]$\lambda$3727 line, in order to obtain a self-consistent way of computing the dust corrected $SFR_{[OII]_{0}}$ from [OII]$\lambda$3727 information alone, without the aid of additional multi-wavelength photometric data.
An anti-correlation ($r_{xy}{=}-0.56$) was found between the rest-frame equivalent width (EW) of the [OII]$\lambda$3727 line (i.e. the blended doublet) and the UV continuum slope (Fig. \ref{airx_oii}), both for galaxies with and without IR data.
A linear fit gives the relation
\begin{equation}
\beta_{phot}~=~(-1.35\pm0.20)~\times~log(EW_{rest})~+~(0.91\pm0.30)
\end{equation}
with an r.m.s. dispersion of the residuals ${\sim}0.6$. 
Since $\beta$ is related to dust attenuation (Eq. 11), the relation with EW implies a relation between EW and dust attenuation for the galaxies in our sample. 
Combining Eq. 11 and Eq. 13 we obtain\footnote[10]{We propose as the relevant relation the one between the EW and $\beta$, instead of the one directly linking the EW to $A_{IRX}$ because $\beta$ is the only dust attenuation related quantity that we can directly measure in all the galaxies in our sample. In the original definition (Eq. 5), $A_{IRX}$ is a directly measured quantity, but it is only available for galaxies with IR data. On the other hand, $A_{IRX}$ derived from $\beta_{phot}$ (Eq. 11) is available for the entire sample, but it is a derived quantity. However, we report that a direct fit of $A_{IRX}$ derived from $\beta_{phot}$ (Eq. 11) and $log(EW_{rest})$ would give a relation consistent with Eq. 14. The same result would be obtained with a direct fit of $A_{IRX}$ as in the original definition (Eq. 5) and $log(EW_{rest})$, for IR-detected galaxies.}
\begin{equation}
A_{IRX}~=~(-1.39\pm0.26)~\times~log(EW_{rest})~+~(4.48\pm0.35),
\end{equation}
Despite all our efforts to clean the sample from galaxies in which older stellar populations might contribute to dust heating (i.e. $L_{IR}$) and continuum reddening, we cannot completely exclude a mild contribution that might increase the scatter in the $\beta_{phot}$ vs. $log(EW_{rest})$ relation in the low-EW$_{[OII]}$ part of the sample, given the presence of Balmer absorption lines (stronger in A-type stars) and CaII $H{\&}K$ lines in the low-EW$_{[OII]}$ stacked spectrum (Fig. \ref{airx_oii}). On the other hand, we measure a stronger EW of the MgII$\lambda$2800 ISM absorption line (i.e. the blended doublet) in the low-EW$_{[OII]}$ stacked spectrum (Fig. \ref{mgii}): ISM absorption lines have been observed to be stronger in SFGs with redder $\beta$ and characterized by higher dust extinction \citep{shapley2003, talia2012}.

\subsection{Continuum vs. nebular attenuation}\label{sec:Continuum vs. nebular attenuation}

We cannot use directly $A_{IRX}$ derived in Eq. 14, i.e. the attenuation towards the continuum, to correct the [OII]$\lambda$3727 luminosity, first because of its dependency on wavelength, and second because of the extra amount of reddening suffered by nebular emission, with respect to the continuum, claimed by many authors starting from \citet{calzetti1997}. 
The first point means that an extinction curve must be assumed. It is known that the effects of dust are stronger at shorter wavelengths, therefore correcting a luminosity measured at $\lambda$3727$\AA$ using an attenuation measured at $\lambda$1500$\AA$ would produce an over-correction.
The extra reddening towards nebular lines is an issue still strongly debated. In fact, though many authors have confirmed the original findings by \citet{calzetti1997, calzetti2000} also at high redshift \citep[see e.g.][]{forsterschreiber2009, wuyts2011}, other studies claim a ratio between continuum and nebular attenuation much closer to unity \citep[]{reddy2010, kashino2013, pannella2014, puglisi2015}. 

To summarize, the dust extinction corrected SFR is
\begin{equation}
SFR_{[OII]_{0}} (M_{\odot}yr^{-1}) = 0.9 \times 10^{-41} L_{[OII]} \times 10^{0.4 A_{[OII]}} (erg s^{-1}),
\end{equation}
where $L_{[OII]}$ is the observed, uncorrected [OII]$\lambda$3727 luminosity.
If we parametrize the extra reddening as
\begin{equation} 
E(B-V)_{neb} = E(B-V)_{cont} / f
\end{equation}
and assume a reddening curve ($\kappa_{\lambda}$), then $A_{[OII]}$ is defined as
\begin{equation}
A_{[OII]} = A_{IRX} \times (\kappa_{H_{\alpha}} / \kappa_{1500 \AA}) \times f^{-1}.
\end{equation}
We use the value of the extinction curve at H$\alpha$ as prescribed by \citet{kennicutt1998} because of the manner in which the [OII]$\lambda$3727 luminosities were calibrated through the [OII]/H$\alpha$ ratio.

In the UV regime a \citet{calzetti2000} law is a fair assumption for the reddening curve, for our sample. In the optical range with our data we are unable to constraint the reddening curve and we decided to assume the same \citet{calzetti2000} law to ease the comparison with recent works on differential attenuation \citep[e.g.][]{kashino2013, wuyts2013, price2014}, though it is still unclear if this is an appropriate assumption in the context of high-redshift galaxies \citep{reddy2015}.

\begin{figure}[h!]
	\centering
	\includegraphics[scale=0.38]{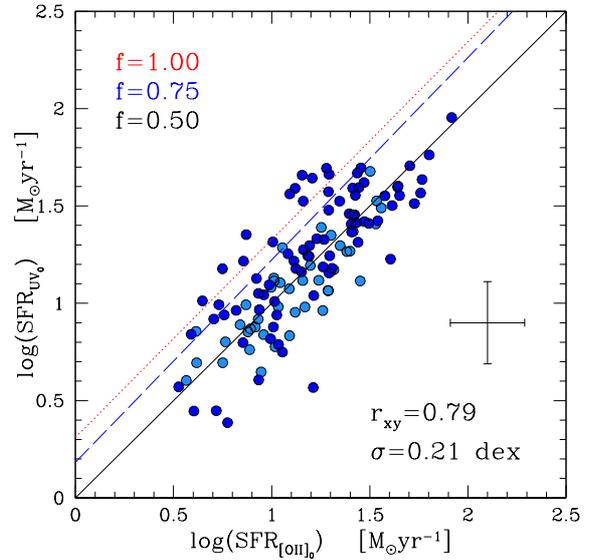}
	\caption{$SFR_{UV_{0}}$ vs. $SFR_{[OII]_{0}}$ corrected for dust extinction using, respectively, $\beta_{phot}$ and Eq. 11, and $EW_{[OII]}$ and Eq. 14 assuming a \citet{calzetti2000} law and $f{=0.50}$. The black line is the 1-to-1 relation that also corresponds also to a linear fit to the points. Blue and red lines show the fits with other two different assumptions of $f$. Blue points are IR-detected galaxies, while light blue points are galaxies with no IR detection.}
	\label{sfruvVSsfroii}
	\end{figure}

We derive dust extinction corrected $SFR_{[OII]_{0}}$ assuming different values of \emph{f} and compare the result to $SFR_{[UV]_{0}}$ (corrected using $\beta_{phot}$ and Eq. 11) to derive an estimate of the best \emph{f}-factor. The best agreement is found with $f{=}0.50$, which is very close to the 0.59 value quantified by \citet{calzetti1997}\footnote[11]{The canonical value quoted by the original works \citep{calzetti1997, calzetti2000} is $f{=}0.44$, while we quote $f{=}0.59$. The difference comes from the fact that in the original papers two different extinction curves for the nebular and continuum emission were used, while we are assuming the same Calzetti law for both emissions \citep{pannella2014, steidel2014}}.
In Fig. \ref{sfruvVSsfroii} where we also plot, for comparison, the fits corresponding to other two different choices of \emph{f} (0.75 and 1.0). The small dispersion (${\sim}0.2$) is due to the correlation of the corrections on the two axes, since we derive $A_{[OII]}$ from $A_{IRX}$. Though we cannot draw any conclusive statement, our data appear to be more consistent with the need of an extra attenuation towards nebular lines in galaxies at high redshift, than with studies that claim an \emph{f} coefficient much closer to unity.

We stress that the relevant relation that we derive in this work is the one linking $\beta_{phot}$ and $EW_{[OII]}$. Then, the actual attenuation of the [OII]$\lambda$3727 line luminosity depends on the $(\kappa_{H_{\alpha}} / \kappa_{1500 \AA}){\times}f^{-1}$ term in Eq. 17, i.e. on the combination of the assumptions about the reddening laws respectively for the stellar continuum and nebular emission, and the best \emph{f} value that we derive by forcing agreement between $SFR_{[OII]_{0}}$ and our reference SFR. With this approach, different assumptions about the reddening law will produce different estimates of \emph{f} \citep{puglisi2015}, but the resulting $SFR_{[OII]_{0}}$ values will be always, by construction, consistent with one another.

As an exercise we derived the \emph{f} value under a different assumption for the nebular attenuation law. We choose a Galactic extinction curve, consistently with the original prescription by \citet{calzetti1997}, in particular the \citet{cardelli1989} one, while for the stellar continuum we continued to assume the \citet{calzetti2000} law \citep[][]{reddy2015, steidel2014}. With this choice of attenuation curves the best agreement between $SFR_{[OII]_{0}}$ and $SFR_{[UV]_{0}}$ is found with $f{=}0.37$.

To conclude this section we summarize the steps to derive the dust extinction corrected $SFR_{[OII]_{0}}$:
\begin{enumerate}
\item Derive the attenuation towards the continuum ($A_{IRX}$) from the rest-frame $EW_{[OII]}$ and its relation with $\beta_{phot}$ (Eq. 13-14).
\item Convert $A_{IRX}$ into $A_{[OII]}$ assuming appropriate reddening curves respectively for the stellar continuum and nebular emission, and an extra-attenuation factor (Eq. 17). 

In this work we assume a \citet{calzetti2000} law for both the stellar continuum and nebular emission and apply an $f$-factor $f{=}0.50$. If a \citet{cardelli1989} Galactic extinction curve were instead applied to nebular emission, while keeping the Calzetti law for the stellar continuum, the $f$-factor to be used would be $f{=}0.37$.
\item Derive $SFR_{[OII]_{0}}$ by correcting the observed [OII]$\lambda$3727 line luminosity with $A_{[OII]}$ (Eq. 15); in this work we assume a \citet{kennicutt1998} calibration.
\end{enumerate}

Finally, we are aware that our relations are taking the calibration of the [OII]$\lambda$3727 line as a SFR indicator as a closed box because with the data in our possess we cannot tackle the issue of how the [OII]/H$\alpha$ ratio varies across our sample, as a function of different gas properties. For this reason our method for deriving the $SFR_{[OII]_{0}}$ should be applied carefully to individual galaxies. However, our calibration will be useful for computing the statistical star formation and extinction properties of large high-redshift galaxy samples, when no other information about nebular extinction is available.

\section{A quick look at SFR from SED fitting}\label{sec:SFR from SED fitting}

We dedicate a brief section to the SFR estimate derived by fitting the galaxy UV-to-NIR broad-band photometry to the spectral energy distribution (SED) of synthetic stellar populations. For this exercise we do not consider the FIR data but use only photometry from U band to IRAC 5.8$\mu$m, to see how the results of our calibrations compare to a different dust-corrected SFR estimate derived without direct information on dust emission.

A widespread approach is to use models in which the star formation history (SFH) is described by an exponentially declining SFR, though in the last years some studies have introduced increasing star formation histories as a better way to describe high-redshift star-forming galaxies, namely at $z{\sim}2$ \citep{maraston2010, reddy2012}. 
We choose the set of models by \citet{maraston2005} and used the \emph{HyperZ} software \citep{bolzonella2000} to perform the SED fitting.
star formation histories were parametrized by exponentials ($e^{-t/\tau}$) with e-folding timescale $\tau$ between 100 Myr and 30 Gyr, plus the case of constant SFR. We assumed a Kroupa IMF, fixed solar metallicity, a Calzetti law for dust extinction, and a minimum age $age_{min}{=}0.09~Gyr$. We derived from simulations the errors on the quantities estimated using SED fitting techniques. We found that errors on both $SFR_{SED}$ and stellar masses are, on average, ${\sim}~30\%$ (Bolzonella, private communication).
	\begin{figure}[h!]
	\centering
	\includegraphics[scale=0.78, trim=40mm 0mm 50mm 0mm, clip=true]{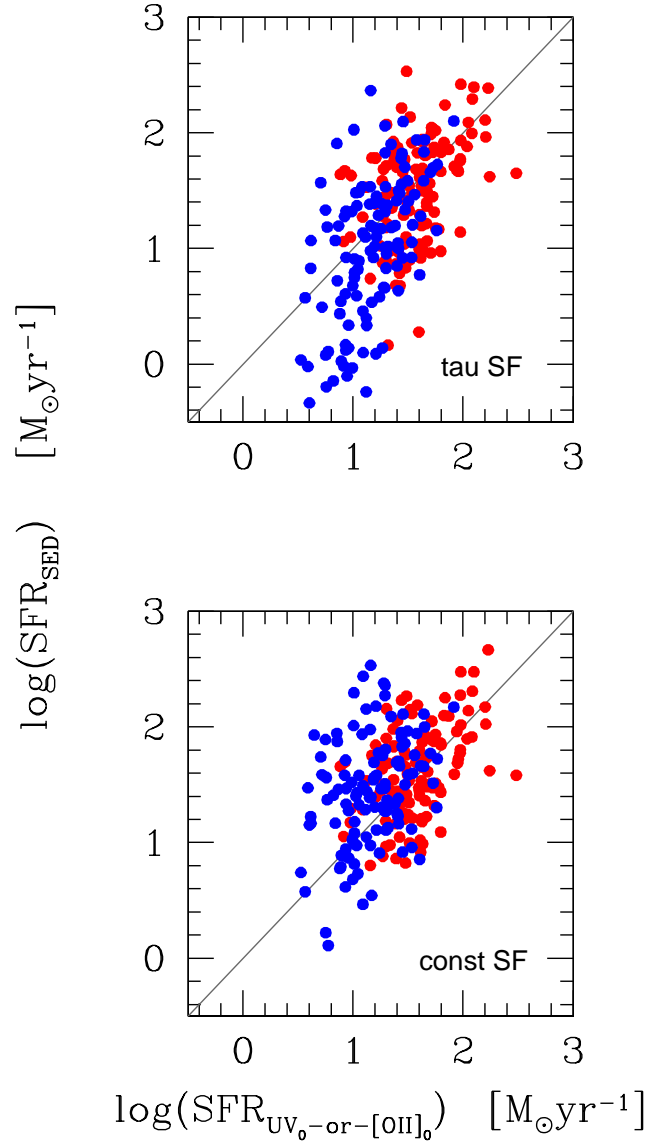}
	\caption{Comparison between $SFR_{SED}$ and SFR estimated from spectroscopic tracers: UV flux, corrected using $\beta$, for galaxies in the \emph{UV sample} (red points), and [OII]$\lambda$3727 flux, corrected using EW$_{[OII]}$, for galaxies in the \emph{[OII] sample} (blue points).}
	\label{sfrspecVSsfrsed}
	\end{figure}

In Fig. \ref{sfrspecVSsfrsed} (top) we compare the results from SED fitting, under the cited assumptions, to the ones previously obtained from our calibrations. In particular, for the \emph{UV sample} we plot $SFR_{UV_{0}}$ corrected using $\beta$, while for the \emph{OII sample} we plot $SFR_{[OII]_{0}}$ corrected as explained at the end of Sec. \ref{sec:Continuum vs. nebular attenuation} (assuming a Calzetti law for both stellar continuum and nebular emission and applying an extra-attenuation $f{=}0.50$).
There is a general fair agreement between the two estimates at high SFR, but in the low-SFR regime the SED estimate is lower than our calibrated values, i.e. lower than $SFR_{IR+UV}$ on which our calibrations are based. This tail at low $SFR_{SED}$ has been reported also by other authors \citep{wuyts2011, reddy2012, arnouts2013, utomo2014}, according to which this difference may have multiple causes. Either IR+UV is overestimating the true SFR due to a non-negligible contribution of old stars to the dust heating or some of the physical assumptions made in the SED fitting procedure do not correctly describe the analysed galaxies, causing $SFR_{SED}$ to underestimate the true SFR. In our case, we tend to exclude the former explanation because we specifically excluded galaxies showing evidence of old stellar populations in their spectra. 
To investigate thoroughly if and how degenerate SED fitting parameters (i.e. SFH, age, attenuation law, metallicity) should be tuned in order to recover the true SFR is beyond the scope of this work. Therefore we limited ourselves to a quick test and we changed only one of the main assumptions, namely the SFH, by forcing all galaxies to a constant SF (and leaving all the other parameters unchanged). We note that, in our sample, when $\tau$ is left as a free parameter only for $\sim20\%$ of the galaxies the best $\chi^{2}$ is obtained with a constant star formation history. The results are shown in Fig. \ref{sfrspecVSsfrsed}, bottom plot.
Assuming a constant SFH the less star-forming galaxies are forced to values of their $SFR_{SED}$ that are in better agreement with our calibrated estimate with respect to the results obtained by leaving $\tau$ a free parameter \citep[see also][]{wuyts2011}, though especially in the $[OII] sample$ $SFR_{SED}$ seems to be systematically slightly higher than our IR-calibrated values.

\section{An application: the SFR vs. Mass relation}\label{sec:SFRvsMASS}

An obvious application of the calibrations derived in the previous section is the study of the SFR vs. stellar mass relation. 

\begin{figure}[h!]
	\centering
	\includegraphics[scale=0.43]{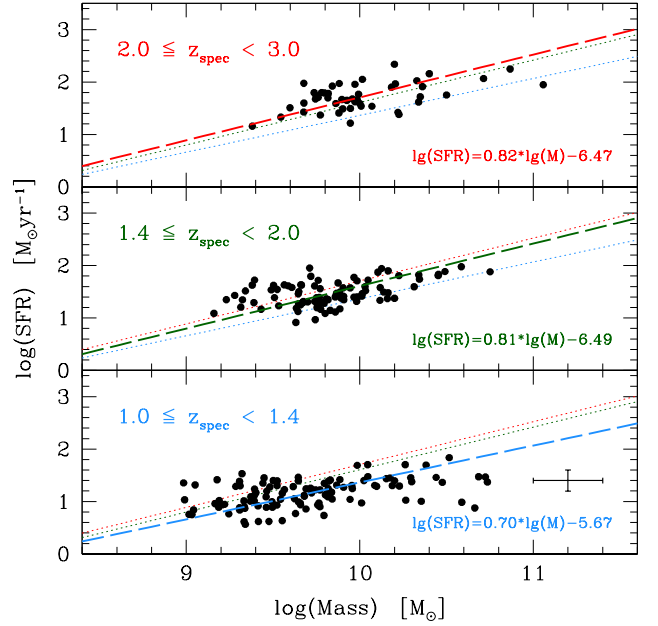}
	\caption{SFR vs. Mass in three redshift bins, each spanning ${\sim}1.2$ Gyrs. For each galaxy, the plotted SFR estimate is $SFR_{IR+UV}$ for IR-detected galaxies and $SFR_{[OII]_{0}}$ or $SFR_{UV_{0}}$ for galaxies with no IR data respectively in the \emph{[OII] sample} and in the \emph{UV sample}. In each panel the three lines are the fits (OLS bisector) to the data in the three redshift bins. Different colours identify each redshift bin.}
	\label{ms}
	\end{figure}

The SFR-M$_{\star}$ relation is expected to evolve with redshift \citep{daddi2007, noeske2007, wuyts2011, whitaker2012, heinis2013, buat2014}, therefore we divided our sample into three redshift bins, each spanning ${\sim}1.2$ Gyrs (Fig. \ref{ms}).
The SFR for each galaxy is chosen as follows: $SFR_{IR+UV}$ for IR-detected galaxies and the appropriate IR-calibrated SFR estimate ($SFR_{[OII]_{0}}$ or $SFR_{UV_{0}}$) for galaxies not detected in IR, depending on the redshift. Stellar masses were computed using \citet{maraston2005} models with the assumptions outlined in the previous section and constant SF.

We find that our galaxies lie on a close linear relation with $\sigma{\sim}0.3$ dex and $r_{xy}{\sim}0.6$ at all redshifts. 
There are weak trends both for the slope and the normalization, with the slope becoming steeper and the normalization higher with increasing redshift, though the completeness limits and spectroscopic requirements of our sample do not allow a robust qualification.
However, we note that the dispersion that we find in the relation is consistent with that often quoted in the literature \citep[see e.g.][]{daddi2007, noeske2007, whitaker2012, buat2014}. Also, the same trend of the normalization with redshift has been reported in previous works, while there is not a consensus yet on the evolution of the slope with redshift \citep{elbaz2007, noeske2007, whitaker2012, heinis2013, buat2014, pannella2014}, as an effect of different sample selections, completeness limits and chosen SFR estimators. 
In the redshift range $1{<}z{<}3$ literature quoted values of the slope range from $\sim$0.7 to $\sim$0.9 \citep[see also][]{wuyts2011, rodighiero2011, kashino2013, rodighiero2014}, with which our findings are consistent.

\section{Conclusions and summary}\label{sec:Summary and discussion}

In this paper we use a sample of galaxies drawn from the GMASS survey to study different SFR estimators. Our aim is to use IR data to derive empirical calibrations to correct UV and [OII]$\lambda 3727$ luminosities for dust extinction.
To do this, we concentrated on a well-controlled spectroscopic sample, rich of ancillary data. 
	\begin{table*}[t!]
	\caption{Summary table of all the relations derived in the paper. For each relation, observed and derived quantities are indicated, as well as slope and intercept of the linear relation, with associated uncertainties. Relations are given in the form: $Derived = Slope \times Observed + Intercept$.}
	\label{tabellona}
	\centering                          
	\begin{tabular}{c c | c c | c | c | l}        
	\hline\hline                 
	Observed & Derived & Slope & Intercept & z & Notes\tablefootmark{a} & Sec.\tablefootmark{b}\\   
	\hline\hline                 
	$\beta_{spec}$ & $A_{IRX}$ & 1.10 $\pm$ 0.23 & 3.33 $\pm$ 0.24 & $1.6{<}z{<}3.0$ & spectral coverage: $1200-2600\AA$ & \ref{sec:The UV continuum slope}, \ref{sec:Appendix}\\
	\hline
	$\beta_{phot}$ & $A_{IRX}$ & 1.45 $\pm$ 0.25 & 3.63 $\pm$ 0.26 & $1.6{<}z{<}3.0$ & & \ref{sec:The UV continuum slope}\\
	\hline
	\emph{(B-z)}   & $A_{IRX}$ & 1.64 $\pm$ 0.13 & 0.77 $\pm$ 0.13 & $1.4{<}z{<}2.5$ & & \ref{sec:Extinction from observed colour}\\
	\hline
	$\beta_{phot}$ & $A_{IRX}$ & 1.03 $\pm$ 0.26 & 3.54 $\pm$ 0.25 & $1.0{<}z{<}1.6$ & & \ref{sec:The OII sample}\\
	\hline
	log($EW_{[OII]}$)\tablefootmark{c} & $\beta_{phot}$ & -1.35 $\pm$ 0.20 & 0.91 $\pm$ 0.30 & $1.0{<}z{<}1.6$ & $[OII]\lambda3727$ emission line  & \ref{sec:The equivalent width vs. beta relation}\\
	\hline
	log($EW_{[OII]}$)\tablefootmark{c} & $A_{IRX}$\tablefootmark{d} & -1.39 $\pm$ 0.26 & 4.48 $\pm$ 0.35 & $1.0{<}z{<}1.6$ & $[OII]\lambda3727$ emission line  & \ref{sec:The equivalent width vs. beta relation}\\
	\hline\hline
	\end{tabular}
	\tablefoot{
	\tablefoottext{a}Spectroscopic requirements, when needed.
	\tablefoottext{b}Reference to the section in the paper where the relation has been derived.
	\tablefoottext{c}Rest-frame. The convention of positive EW for emission lines is adopted.
	\tablefoottext{d}$A_{IRX}$ is the attenuation towards the stellar continuum at $\lambda{=}1500\AA$. The attenuation for the [OII]$\lambda$3727 line is defined as $A_{[OII]}{=}A_{IRX}{\times}(\kappa_{H_{\alpha}}{/}\kappa_{1500 \AA}){\times}f^{-1}$, where $\kappa_{\lambda}$ is the reddening curve. Assuming a Calzetti law for both the stellar continuum and nebular emission we estimate $f{=}0.50$. Assuming instead a \citet{cardelli1989} law for nebular emission and Calzetti law for the stellar continuum we estimate $f{=}0.37$ (Sec. \ref{sec:Continuum vs. nebular attenuation}).
	 }
	\end{table*}

We started by selecting a sample of SFGs with a spectroscopic redshift $1{<}z{<}3$ from a pure magnitude-limited parent sample ($\emph{m}_{4.5}{<}23.0$). In the chosen redshift range this selection is most sensitive to stellar mass. In particular, the limiting mass sensitivities are $log(M/M_{\odot}) \sim$ 9.8, 10.1, and 10.5 for $z =$ 1.4, 2, and 3, respectively. 
Our sample can be divided into two sub-samples with homogeneous rest-frame wavelength coverage: 
\begin{itemize}
\item	galaxies at $1{<}z{<}1.6$ that cover the range $\sim2700-4300\AA$ (\emph{[OII] sample}), where the [OII]$\lambda$3727 emission line can be observed; 
\item	and galaxies at $1.6{<}z{<}3$ that cover the range $\sim1100-2800\AA$ (\emph{UV sample}), in whose spectra strong ISM absorption lines can be detected.
\end{itemize}

We excluded from the sample all quiescent galaxies and AGNs, identified on the basis of spectroscopic features and X-ray luminosity. We also used three continuum indices (MgUV, C(29-33), D4000) to clean the resulting preliminary selection of SFGs from galaxies showing some evidence of the presence of old stellar populations that could bias the correct interpretation of the dust- and SFR-related observables. 
The final sample includes 286 SFGs, of which on third has IR information coming from Spitzer-MIPS and Herschel-PACS, and another one third has only a Spitzer-MIPS detection. 

Though inevitably not complete, especially due to the spectroscopic requirement that introduces a bias towards brighter and bluer objects, our sample is fairly representative of a specific galaxy population whose properties may be summarized as follows:
\begin{itemize}
\item redshift between $z\sim1$ and $z\sim3$;
\item intermediate stellar mass (approximately in the range $10^{9.2}{<}M/M_{\odot}{<}10^{10.2}$);
\item blue rest-frame colours: $(NUV-r){<}3 \ \& \  (r-K){<}1.4$ or \\ $(U-V){<}1.2 \ \& \ (V-J){<}1.5$;
\end{itemize}

The bolometric IR luminosity, $L_{IR}$, was derived for all IR-detected galaxies, using the most up-to-date models, i.e. SED fitting to U-to-FIR broad-band photometry with Magphys \citep{dacunha2008} for PACS-detected galaxies, and 24$\mu$m-to-bolometric $L_{IR}$ correction using main sequence SED models by \citet{magdis2012} for MIPS-detected galaxies with no PACS data. $L_{IR}$ values were then converted to SFR, to which we added a second component accounting for unobscured SFR, in order to recover the total SFR ($SFR_{IR+UV}$). 
Assuming $SFR_{IR+UV}$ as our benchmark SFR estimate, we derived some relations to correct UV and [OII]$\lambda 3727$ luminosities for dust extinction and thus obtain a set of consistently calibrated SFR estimators covering a wide range of redshifts and SFR regimes that will be particularly useful in view of the large spectroscopic surveys that are currently on-going or will be carried out in the near future (for example, VUDS \citep{lefevre2014}, VANDELS, BigBOSS \citep{schlegel2012}, Euclid \citep{laureijs2011}, WFIRST \citep{spergel2013}).

In Table \ref{tabellona} we summarize all the relations between attenuation and different spectral properties and colours that were derived in the paper. 
In this respect, the main results of our analysis can be summarized as follows:

\begin{enumerate}
\item The UV continuum slope was derived, on spectra for the galaxies in the \emph{UV sample} and from photometry for galaxies in the \emph{[OII] sample}. The relation between dust attenuation ($A_{IRX}$) and $\beta$ was studied and found to be broadly consistent with literature results at the same redshift, though with a larger dispersion with respect to UV-selected samples, which we ascribe to our parent sample being mass-selected. Our results also suggest that the Calzetti law is valid up to $z{\sim}3$, though it tends to over-predict dust attenuation for galaxies with the reddest slope.

\item Using photometric information, the \citet{daddi2004} relation between (B-z) colour and attenuation in galaxies at $1.4{<}z{<}2.5$ was updated using $A_{IRX}$ as an independent dust attenuation tracer. 

\item In the \emph{[OII] sample} ($1{<}z{<}1.6$), where the [OII]$\lambda$3727 emission line can be detected, we found an anti-correlation between the rest-frame EW of the line and $\beta_{phot}$, which is the main result in the paper. Since $\beta$ is related to dust attenuation, this relation implies a relation between $EW_{[OII]}$ and dust attenuation for the galaxies in our sample. 

\item We tested the issue of differential attenuation towards, respectively, stellar continuum and nebular emission. Though we cannot draw any conclusive statement, our results are in line with the traditional prescription of extra attenuation towards nebular lines. 

\item Calibrated SFRs were used to discriminate between different SFHs in a quick comparison to SFR estimates from SED fitting to UV-to-NIR broad-band photometry (i.e. without including FIR information), and the best agreement was found with constant star formation, as opposed to exponentially declining. 

\item Finally, we divided our sample into three redshift bins and using the calibrations in Table \ref{tabellona} we studied the relation between SFR and stellar mass. The galaxies in our sample lie on a quite close linear relation ($\sigma{\sim}0.3$ dex, $r_{xy}{\sim}0.6$) at all redshifts, with a slope ${\sim}0.7-0.8$, consistently with other literature results in the same redshift range. We also find weak trends both for the slope and the normalization, with the slope becoming steeper and the normalization higher with increasing redshift, though the completeness limits and spectroscopic requirements of our sample do not allow a robust qualification.
\end{enumerate} 


\begin{acknowledgements}
We acknowledge the grants ASI n.I/023/12/0 "Attivit\`a relative alla fase B2/C
per la missione Euclid" and MIUR PRIN 2010-2011 "The dark Universe and the
cosmic evolution of baryons: from current surveys to Euclid".
We acknowledge the use of the software IRAF \citep{tody1993} in the spectroscopic analysis.
MT would like to thank Gianni Zamorani for illuminating discussions on various aspects of the paper. 
The authors also thank the anonymous referee for very useful comments that helped to improve the clarity of the paper.

\end{acknowledgements}


\bibliographystyle{aa} 
\bibliography{references} 

\appendix
\section{Derivation of the UV continuum slope from spectra}\label{sec:Appendix}

	\begin{figure}[h!]
	\centering
	\includegraphics[scale=0.83, trim=0mm 0mm 100mm 0mm]{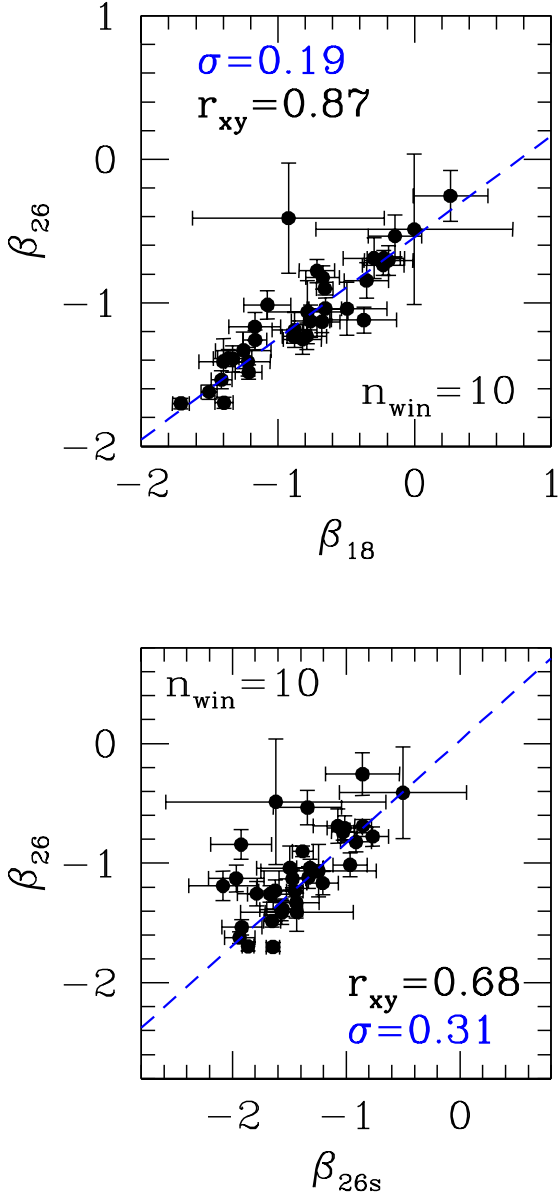}
	\caption{Comparison between $\beta_{26}$, $\beta_{18}$, and $\beta_{26s}$, for the galaxies whose spectrum covers the wavelength range $1250{-}2600{\AA}$. See the text for more details about the definitions of $\beta$. Blue dashed lines indicate the linear fits to the data.}
	\label{uvslope_vs_uvslope}
	\end{figure}

In Section \ref{sec:The UV continuum slope} we studied the relation between the UV continuum slope and dust attenuation. 
In galaxies dominated by a young stellar population, the shape of the UV continuum can be fairly accurately approximated by a power law $F_\lambda{\varpropto}\lambda^\beta$, where $F_\lambda$ is the observed flux ($erg~s^{-1}~cm^{-2}~\AA^{-1}$) and $\beta$ is the continuum slope \citep{calzetti1994}. The wavelength range in which $\beta$ is defined goes from ${\sim}1250{\AA}$ to ${\sim}2600{\AA}$. In this range, \citet{calzetti1994} defined ten windows to be used to measure the continuum, chosen to avoid the strongest spectral features. 
For galaxies at $z{\geq}1.6$ $\beta$ was derived through an error-weighted OLS $(Y{\mid}X)$, by fitting the average fluxes in the ten windows defined by \citet{calzetti1994}, in the $log(F_\lambda)~-~log(\lambda)$ plane. More details about the procedure can be found also in \citet{talia2012}. We define the slope computed using all the ten windows as $\beta_{26}$. Since not all spectra cover the entire $1250{-}2600{\AA}$ range, we define also two shorter versions of $\beta$: $\beta_{18}$, ranging from ${\sim}1250{\AA}$ to ${\sim}1800{\AA}$, and $\beta_{26s}$, ranging from ${\sim}1550{\AA}$ to ${\sim}2600{\AA}$. With respect to the ten original windows, $\beta_{18}$ is measured using the 7 bluer windows (1st to 7th), while $\beta_{26s}$ is measured using the 6 redder ones (5th to 10th). 
 
In Table \ref{slope} we indicate the number of galaxies for which each version of the slope could be computed, depending on the wavelength range covered by the spectra.

In Fig. \ref{uvslope_vs_uvslope} we compare the three definitions of $\beta$ for the galaxies whose spectrum covers the entire wavelength range $1250{-}2600{\AA}$. $\beta_{18}$ is always redder (more positive) than $\beta_{26}$, and the discrepancy increases for increasing value of the slopes \citep{calzetti2001}.  

	\begin{table}[h!]
	\caption{Number of galaxies for which the different definitions of $\beta$ could be computed.}
	\label{slope}
	\centering                          
	\begin{tabular}{l | c c c}        
	\hline\hline  
	Windows				&	1st-10th	&	1st-7th 	&	5th-10th	\\
	\hline	
	Tot. no. galaxies		&	35		&	33		&	46		\\
	\hline	
	only \emph{24$\mu$m}	&	14		&	11		&	17		\\
	\hline	
	\emph{24$\mu$m} and \emph{PACS}		&	8		&	1		&	11		\\
	\hline\hline
	\end{tabular}
	\end{table}

There is a strong correlation between the two definitions of the slope. In particular, applying an error-weighted linear fit we obtain the following relation ($r_{xy}{\sim}0.9$):
\begin{equation}
\beta_{26}~=~(0.71\pm0.03)\times\beta_{18}+(-0.54\pm0.03)
\end{equation}

On the other hand, $\beta_{26s}$ is systematically slightly bluer (more negative) than $\beta_{26}$, which is quite intuitive, since $\beta_{18}$ and $\beta_{26s}$ basically sample the two halves of $\beta_{26}$. In this case, a linear fit gives:
\begin{equation}
\beta_{26}~=~(0.86\pm0.04)\times\beta_{26s}+(0.03\pm0.06)
\end{equation}
with a slightly larger dispersion than in the previous case and no dependence on $\beta_{26}$.
The likely explanation of the discrepancy between the three definitions is the presence of a large number of closely spaced Fe absorption lines in the $2300-2800\AA$ range \citep{leitherer1999, calzetti2001}. 
For each spectrum in the \emph{UV sample} the appropriate definition of the slope was computed, depending on the wavelength coverage. Then all the $\beta_{18}$ and $\beta_{26s}$ values were scaled to $\beta_{26}$ using Eq. A.1 and Eq. A.2: these homogenised values are the ones that were plotted in Fig. \ref{slope_vs_irx} and used to derive the relation given in Sec. \ref{sec:The UV continuum slope}.

As explained in Sec. \ref{sec:The UV continuum slope}, $\beta$ can be derived also from photometry. In the \emph{[OII]-sample} the available photometry forced us to adopt a short wavelength baseline and we defined $\beta_{phot}$ in the range $\lambda_{rest}{\sim}1500{-}2600\AA$. As we did with the spectroscopic determination of the slope, we used the galaxies in the \emph{UV-sample} to test the effect of different wavelength baselines also in the computation of the slope from photometry. We found that the values of $\beta_{phot}$ derived using, respectively, the total baseline ($1250-2600\AA$) and a shorter one ($1500-2600\AA$) are broadly consistent with one another, with the short-based values being on average slightly bluer than the long-based ones, though to a lesser extent than in the case of $\beta_{spec}$. Since the short wavelength baseline is available from photometry for the entire galaxy sample, we used directly the values of $\beta_{phot}$ in the range $\lambda_{rest}{\sim}1500-2600\AA$ to obtain the relations presented in Sec. \ref{sec:The UV continuum slope} and \ref{sec:The OII sample}, without the need to scale them to the total wavelength baseline.

%


\end{document}